\PassOptionsToPackage{table,xcdraw}{xcolor}
\documentclass[sigconf,balance=false]{acmart}
\usepackage{adjustbox}
\usepackage{popets}
\usepackage{tikz}
\usepackage{float}
\usepackage{makecell}
\setcopyright{popets}
\copyrightyear{2024}

\acmYear{2024}
\acmVolume{2024}
\acmNumber{4}
\acmDOI{10.56553/popets-2024-0106}
\acmISBN{}
\acmConference{Proceedings on Privacy Enhancing Technologies}
\newcommand{\sectionref}[1]{$\S$\ref{#1}}

\begin{document}

\title[Automatic Generation of Web Censorship Probe Lists]{Automatic Generation of Web Censorship Probe Lists}


\author{Jenny Tang}
\affiliation{%
  \institution{Carnegie Mellon University}
  \city{}
  \country{}
  }
\email{jennytang@cmu.edu}
\author{L\'{e}o Alvarez}
\affiliation{%
  \institution{EPFL and CMU}
  \city{}
  \country{}
  }
\email{leo.alvarez@alumni.epfl.ch}
\author{Arjun Brar}
\affiliation{%
  \institution{Carnegie Mellon University}
  \city{}
  \country{}
  }
\email{abrar@cmu.edu}
\author{Nguyen Phong Hoang}
\affiliation{%
  \institution{University of British Columbia}
  \city{}
  \country{}
  }
\email{nphoang@cs.ubc.ca}
\author{Nicolas Christin}
\affiliation{%
  \institution{Carnegie Mellon University}
  \city{}
  \country{}
  }
\email{nicolasc@cmu.edu}

\renewcommand{\shortauthors}{Tang et al.}

\begin{abstract}
  Domain probe lists---used to determine which URLs to probe for Web
censorship---play a critical role in Internet censorship measurement
studies. Indeed, the size and accuracy of the domain probe list limits
the set of censored pages that can be detected; inaccurate lists can
lead to an incomplete view of the censorship landscape or biased
results. Previous efforts to generate domain probe lists
have been mostly manual or crowdsourced. This approach is
time-consuming, prone to errors, and does not scale well to the
ever-changing censorship landscape.

In this paper, we explore methods for automatically generating
probe lists that are both comprehensive and up-to-date for Web
censorship measurement. We start from an initial set of 139,957 unique
URLs from various existing test lists consisting of pages from a variety
of languages to generate new candidate pages. By analyzing content from
these URLs (i.e., performing topic and keyword extraction), expanding these topics, and using
them as a feed to search engines, our method produces 119,255 new URLs
across 35,147 domains. We then test the new candidate pages
by attempting to access each URL from servers in eleven different global
locations over a span of four months to check for their connectivity and
potential signs of censorship. Our measurements reveal that our method
discovered over 1,400 domains---not present in the original dataset---we
suspect to be blocked. In short, automatically updating probe lists is
possible, and can help further automate censorship measurements at scale.

\end{abstract}


\maketitle

\section{Introduction}
The Web, despite its facade of open and unrestricted access, is subject to
various forms of control and censorship. As extensively documented over the
years~\cite{Deibert2010AccessC, deibert2020reset}, many
entities---including governments~\cite{Aryan:2013, Gebhart2017InternetCI,
Ramesh2020DecentralizedCA, USESEC21:GFWatch, Nourin2023:TMC} and private
interests~\cite{Deibert2022SubversionIT, Ruan2020TheIO}---control the free
flow of information and knowledge in different
ways~\cite{Gill2015CharacterizingWC}, ranging from blocking access to web
pages via poisoning DNS resolutions~\cite{anon2012cn.collateral,
Farnan2016a, Pearce:2017:Iris, USESEC21:GFWatch, Hoang2022a} and TCP/IP
packet filtering~\cite{Clayton2006IgnoringTG, Park2010EmpiricalSO,
Wang2017YourSI, Hoang2019:I2PCensorMeasure, Wu2023HowTG, Hoang2024:GFWeb},
to injecting block pages~\cite{jones.2014.blockpages,
Hoang2019:I2PCensorMeasure}, to removing information from the
Internet~\cite{knockel2018can, streisand2023have}, or spreading competing
(false) narratives~\cite{hulcoop2017tainted, lim2019burned,
Hanley2023SpeciousST}, among others. This work focuses on the detection of
Internet censorship, specifically the blocking of web pages, which presents
both ethical and technical challenges.

Identifying blocked pages is not straightforward due to the massive scale
of the Internet and the dynamic nature of online content. As of 2023, there
are over 359 million domain name registrations~\cite{verisign_tlds}, making
comprehensive daily monitoring across all locations impractical. Current
approaches, including crowdsourced probe lists, are invaluable but come with
limitations. They often reflect the biases and regional focus of
contributors, potentially missing censorship of certain topics while
requiring significant manual effort to maintain. 

Some examples include the Berkman Klein Center, which used to maintain a list of URLs
intended to estimate the (in)accessibility of different types of content
from different countries~\cite{berkman-testlist}. The Citizen Lab provides
test lists~\cite{CitizenLabList} curated by regional volunteers to be
relevant to specific countries/regions. The Citizen Lab test list is
widely used by several global censorship measurement projects, including
the Open Observatory of Network Interference (OONI)~\cite{OONI-paper}, the
Information Controls Lab (ICLab)~\cite{ICLab-paper}, and Censored
Planet~\cite{CensoredPlanet-paper}.

In this work, we aim to automate the process of updating probe lists, to
better keep pace with the ever-evolving landscape of online censorship.
Despite efforts to maintain and continuously update existing lists, they
require large amounts of effort by human volunteers and researchers.
Sometimes, well-meaning users may contribute websites that they erroneously
believe as being censored, but that are in fact inaccessible due to human error
or transient network issues. Removing these contributions would take
further resources that may be better spent on other tasks. Furthermore, as
news cycles continue and change, what may be blocked one day may be
accessible the next. Therefore, if probe lists are not continuously
updated, they may become outdated and no longer provide useful information,
or worse, provide misleading information and take away resources from
measuring other potentially blocked pages. 

Building on existing lists, we develop a method to discover new URLs and
domains that are potentially blocked, without needing manual curation,
which requires time, effort, and poses potential risks. This not only
allows us to keep up to date with global trends on Internet censorship with
automatically updated lists, it also reduces possible risk to human
volunteers, by relying less on actual people probing potentially sensitive
information.

We take an original set (drawn from publicly available
sources~\cite{weinberg2017topics, WikileaksList, CitizenLabList}) of
139,957 unique URLs from 106,878 unique domains as our starting point. We
will subsequently refer to these initial lists of pages as our ``source
list.'' Unsurprisingly, given how old some of the components of the source list
are, many pages appear to be down, i.e., they can never be reached from any
location, including places (such as the United States) not known for censoring
much data. We thus filter these down to 51,313 live URLs, and use
these to generate a probe list containing 119,255 candidate URLs for
testing~(\sectionref{sec:sourcelist}).

Specifically, using the URLs from the source list, we devise a new method
of generating potentially censored URLs~(\sectionref{sec:llm-pipeline}).
While the high level idea is simple---extract topics and keywords from
pages that were censored, and feed them into a search engine to get more
examples to test for---its implementation is complex, and
requires us to rely on a rather complicated natural language processing
pipeline, due to the impossibility of making any assumption on the content of
the pages being censored (e.g., they may not be in English, they may
include boilerplate that needs to be removed prior to processing, etc.)
That pipeline includes language identification, page tokenization and
translation, and topic and keyword assignment. To avoid snowballing biases,
we further conduct topic expansion using large language models and Google Trends,
and eventually, use web search on this expanded set of topics to discover potentially blocked pages.

The resulting 119,255 unique URLs fall across 35,147 pay-level domains.\footnote{We determine pay-level domains (PLDs)
using the public suffix list~\cite{publicsuffix}.}
Of these, 71,960
URLs from 32,543 domains did not appear in our source lists (even as dead
links). Therefore, the majority of domains our system outputs are new,
though these domains may not be censored.

To rigorously evaluate our generated probe list, we conducted systematic
testing across 11 strategically chosen locations spanning North America,
Europe, and Asia over a five-month period between November 2023 and March
2024~(\sectionref{sec:testing_probe_list}). From each vantage point, we
performed up to 50 iterations of testing, recording the number of URLs that were
accessible, inaccessible, or returned errors. These results were then
compared against a baseline aggregated across five vantage points with high
Internet freedom scores, which we considered our expected benchmark for
accessibility and failures. To obtain more comprehensive and granular
insights, we further validated our findings using OONI 
Probe~(\sectionref{sec:ooni-tests}), renowned for its robust censorship
detection capabilities.

Compared to traditional crowdsourcing approaches, our automated system
demonstrates the ability to generate new URLs and domains at lower costs
and no risks to end users. Notably, our method enabled the discovery of
pages from domains that were not present in our original source lists,
expanding the scope of our investigation beyond known censored
content~(\sectionref{sec:probe_list_eval}).
A particularly compelling finding emerged in our analysis of China. Our
system is highly effective at identifying new domains that are blocked in this
country, with over 1,000 previously undiscovered domains---absent from our
source lists---consistently inaccessible from these locations. This
discovery showcases the efficacy of our automated approach in detecting
censorship patterns that may have been overlooked by traditional manual
curation processes.

Furthermore, our analysis revealed differences between locations in the
blocking methods employed to restrict access to potentially sensitive
pages. Notably, together with DNS and HTTP blocking, our results also
indicated that IP-based blocking also impacts a number of
domains in China. Concurrently, we also identified regional similarities in
the pages that are censored across certain geographies.

\section{Background and Motivation}
\label{sec:background}
As of 2024, the World Wide Web accommodates billions of users,
facilitating the sharing and access of information across 1.09 billion
websites~\cite{WebStats}. However, only about 18\% of these websites,
equating to 200 million, are actively maintained and frequented. On
average, a new website is created and goes live approximately every 3
seconds. Testing the status of a website, especially to determine if
it is censored, involves attempting to access the site by sending a
request to the server hosting the content from a specific location,
and then analyzing the resulting response and behavior. Considering
the impracticality of testing the entire Web, it is crucial to develop
a strategy to effectively narrow down the scope of the search. In
this section, we give an overview of the existing efforts on Internet
censorship measurement, the challenges they face, and how they have
motivated us to conduct this study.

\subsection{Global Censorship Measurement Platforms}
To shed light on the state of Internet censorship, several projects have
been launched to measure and analyze Web accessibility, including
OONI~\cite{OONI-paper}, ICLab~\cite{ICLab-paper}, and Censored
Planet~\cite{CensoredPlanet-paper}.

Relying on a network of volunteers, OONI~\cite{OONI-paper} operates through
user-installed probe software, gathering data on users' web access
attempts. Their analysis focuses on identifying likely censored links. To
eliminate risks associated with volunteer-based measurements,
ICLab~\cite{ICLab-paper} relies on commercial Virtual Private Network (VPN)
services to measure connectivity disruptions. With a different approach,
Censored Planet~\cite{CensoredPlanet-paper} employs various remote
measurement techniques to infer network censorship by making use of public
servers such as open DNS resolvers and echo servers. 

\subsection{Probe List Curation}

Despite their differences in measurement techniques and vantage points, all
three platforms share a common reliance on the Citizen Lab's test
lists~\cite{CitizenLabList} as the primary input into their measurement
pipelines to identify potential censored websites. This is because it is
infeasible to test every website on the Web with adequate frequency.
Therefore, to narrow down the scope of censorship detection, the Citizen
Lab test lists are manually curated by volunteers with some local knowledge
of what websites are prone to censorship or have already been confirmed 
censored.

In addition, there have been also other efforts to build probe lists, such
as the OpenNet Initiative (ONI)~\cite{ONI} and Herdict by the Berkman Klein
Center~\cite{berkman-testlist}. Unfortunately, these projects are no longer
active at the time of writing.

\subsection{Probe List Generation}

Prior efforts such as FilteredWeb~\cite{Darer2017FilteredWebAF} and Hounsel
et al.~\cite{autoblocklist} have also explored approaches to generate probe
lists. These works discover blocked URLs through automated methods of analyzing the
contents of web pages and finding pages with similar topics by using
these topics and keywords as inputs to search engines.

However, our approach goes beyond finding only topics associated with
previously censored pages by extending the search space of potentially censored
topics by leveraging Google Trends and language models like GPT. This
allows us to detect pages that may not be directly related to the
content of existing block lists, enabling the discovery of other similar
and potentially sensitive topics.

Furthermore, while both FilteredWeb and Hounsel et al.'s work focused
regionally on China, our study attempts to provide a more global
perspective on Internet censorship. We additionally conduct repeated measurements
over time and utilizing the OONI Probe~\cite{ooni-cli}, a widely used
censorship detection tool by the anti-censorship community. Thus, our evaluation
of the generated probe list enables more fine-grained detection of censorship
due to different blocking mechanisms, including DNS, TCP/IP, and HTTP. This
comprehensive approach provides higher confidence that the identified
blocked domains are indeed censored, rather than experiencing transient
errors.

\subsection{Motivation}

These observations show that prior initiatives in constructing probe lists
often heavily rely on individual contributions and manual efforts, or are specific to certain regions. These
curated lists also can become outdated quickly and are not refreshed at a
pace matching the dynamic nature of web content and changes in censorship.
In fact, an investigation by Weinberg et al.~\cite{weinberg2017topics}
found that websites hosting sensitive content are often short-lived,
indicating the dynamic and volatile nature of these lists. This volatility
raises questions about the completeness and reliability of the data over
time.

These previous challenges have motivated us to try to
bridge this gap by devising a method to generate a probe list with broad
thematic coverage while minimizing manual efforts to maintain the list.
Updating probe lists is non-trivial, and currently
requires large quantities of manual hours and volunteers. We explore ways to
use language models to reduce some of the onus on human volunteers and
researchers, while maintaining the ability to discover relevant pages
with greater likelihood to be censored. Ultimately, we aim to provide
censorship monitoring tools with a solution to generate probe lists and
narrow the entire web to a selection of websites that are more likely to
be targeted by censoring systems.

\section{Generating the Probe List}
We now document how 
we build our original source list, before describing how we use this
list to produce more candidate pages, i.e., our new probe list. 

\subsection{Source List and Input Sanitization}
\label{sec:sourcelist}

\begin{table}
	\caption{\label{tab:sourcelist} \textbf{Composition of the source list.}}
  \begin{adjustbox}{width=\columnwidth, center}
    \begin{tabular}{lllr}
      \toprule
      \textbf{Group} & \textbf{List name} & \textbf{Source} & \textbf{Number of URLs}\\
      \midrule
      \textit{BlackPink} & Australia 2009 & Wikileaks & 1,168\\
      \textit{BlackPink} & Denmark2008 & Wikileaks &  3,862\\
      \textit{BlackPink} & Finland2009 & Wikileaks &  797\\
      \textit{BlackPink} & Norway2009 & Wikileaks &   3,517\\
      \textit{BlackPink} & Thailand2007 & Wikileaks &  13,428\\
      \textit{BlackPink} & Thailand2008 & Wikileaks &  1,309\\
      \textit{BlackPink} & Thailand2009 & Wikileaks & 398\\
      \textit{BlackPink} & UK2015 & Weinberg et al.~\cite{weinberg2017topics} & 87,598\\
      \textit{CitizenLab} & CitizenLab & Citizen Lab~\cite{CitizenLabList} & 37,570\\
      \midrule
      \multicolumn{3}{c}{Total (with duplicates)} & 150,005\\
      \multicolumn{3}{c}{\textbf{Total (unique)}} & 139,957\\
      \bottomrule
    \end{tabular} 
  \end{adjustbox}
\end{table} 

We begin with a set of known blocked pages, sourced from lists used in
prior work by Weinberg et al.~\cite{weinberg2017topics}, the Wikileaks'
Internet Censorship page~\cite{WikileaksList}, and the Citizen Lab
test lists~\cite{CitizenLabList}. Table~\ref{tab:sourcelist}
details the exact composition of these lists. Most of the lists are
static (i.e., not updated regularly), old, and---as noted by
Weinberg et al.---already contained a lot of broken links in
2017~\cite{weinberg2017topics}. We expect these lists to be even more
stale in 2023. Nonetheless, they can still serve to
seed our probe list generation. 
Weinberg et al.~\cite{weinberg2017topics} distinguish between lists
that contain mostly pornographic material (``pinklists'') and those that
do not (``blacklists''). In the present study, this distinction is less
relevant, and we combine these corpora under a common ``BlackPink''
header,\footnote{Not to be construed as a hidden K-pop
reference.} primarily to distinguish them from the Citizen Lab list
\cite{CitizenLabList}, which is actively maintained and updated
on a regular basis by the OONI community and the Citizen Lab. We
fetched our data from the list on March 31, 2023, from both global
and country-specific test lists, totaling 37,750 URLs.

We combine all the lists and remove duplicate entries, resulting in a
total of 139,957 unique URLs which form the basis of our generations of
new probe list. We will refer to this corpus as the ``source list,'' i.e.,
the input to our probe list generation pipeline.

\noindent\textbf{Removing dead pages.} 
Given the crowdsourced nature of some of our data, we use a multi-step process to eliminate URLs
that are inactive or composed of irrelevant information (e.g., parked
domains). We issue a request to each URL and classify it as dead if the
response:

\begin{itemize}

\item results in invalid redirections (invalid URL formats, infinite or
excessive redirections, redirection to a domain identified as a domain
seller, redirection to a domain in a manually handcrafted list as
suspicious).

\item returns a 4XX error code that was not among 4\{03, 04, 05, 06, 08,
  12, 14, 15, 23, 29\}.

\item returns a 5XX error code that was not among 5\{00, 01, 02, 03, 04,
05, 08, 11, 20, 91\}.

\end{itemize}

\noindent\textbf{Removing content-free pages.} 
We are left with a set of 84,451 URLs which we then test further
with Selenium WebDriver to determine whether the corresponding pages
are active and display meaningful content. To do so, we send HEAD
requests to each of these pages and extract the page content using the
Trafilatura library for analysis~\cite{trafilatura}. Trafilatura allows
us in particular to extract the main body of a page, removing recurring
elements such as menu bars, links, blog rolls, social media buttons, etc.

We then use regular expressions on this main body to identify pages that
do not contain actual content and remove the corresponding URLs from our
list. Examples include YouTube pages with missing or taken down videos,
or error pages. While these remain accessible, the content of these
pages does not provide us any useful information for our pipeline to
generate candidate test pages. Following Weinberg et al.~\cite{weinberg2017topics}'s lead, we also use regular expressions developed
by Szurdi et al. to remove parked domains~\cite{szurdi2014long}.
This step results in 18,911 URLs being removed from the source list,
leaving us with 65,540 URLs for further analysis.

Finally, we test whether the page contains at least
300~characters of text for liveness and language detection. 
Pages with fewer than 300~characters often
contain minimal or no meaningful content at all, and are likely to be
inactive. It is very rare that websites have no text at all, as even
web pages that are characterized by images and videos often contain text
in the form of video descriptions, comments, and even alt-tags.
Furthermore, extracting semantic meaning (which we need to do in 
the next stages of our pipeline) of extremely short text snippets is
challenging as such short snippets are also unlikely to contain enough
information to accurately identify the language of the page. In short,
pages with fewer than 300 characters\footnote{We arrived at 300 characters
from the original length of a tweet---280 characters---that we rounded
up to account for the lack of URL shortening.} often
do not contain enough information to be useful for our pipeline.

This multi-step process results in a final set of 51,313 URLs that we feed
into our pipeline to generate new candidate pages. The small
number of URLs (36.66\% of our original source list) successfully
processed underscores the amount of manual effort required to maintain
probe lists and the importance of keeping them up to date and relevant.

\subsection{Expanding the source list into the probe list}
\label{sec:llm-pipeline}

\usetikzlibrary{shapes.geometric, arrows, calc} 

\tikzstyle{startstop} = [
rectangle, 
rounded corners, 
minimum width=2cm, 
minimum height=1cm,
text centered, 
draw=black, 
fill=red!30,
]

\tikzstyle{process} = [
rectangle, 
minimum width=1.5cm, 
minimum height=1cm, 
text centered, 
text width=1.5cm, 
draw=black, 
fill=orange!30,
]

\tikzstyle{arrow} = [
thick,
->,
>=stealth]

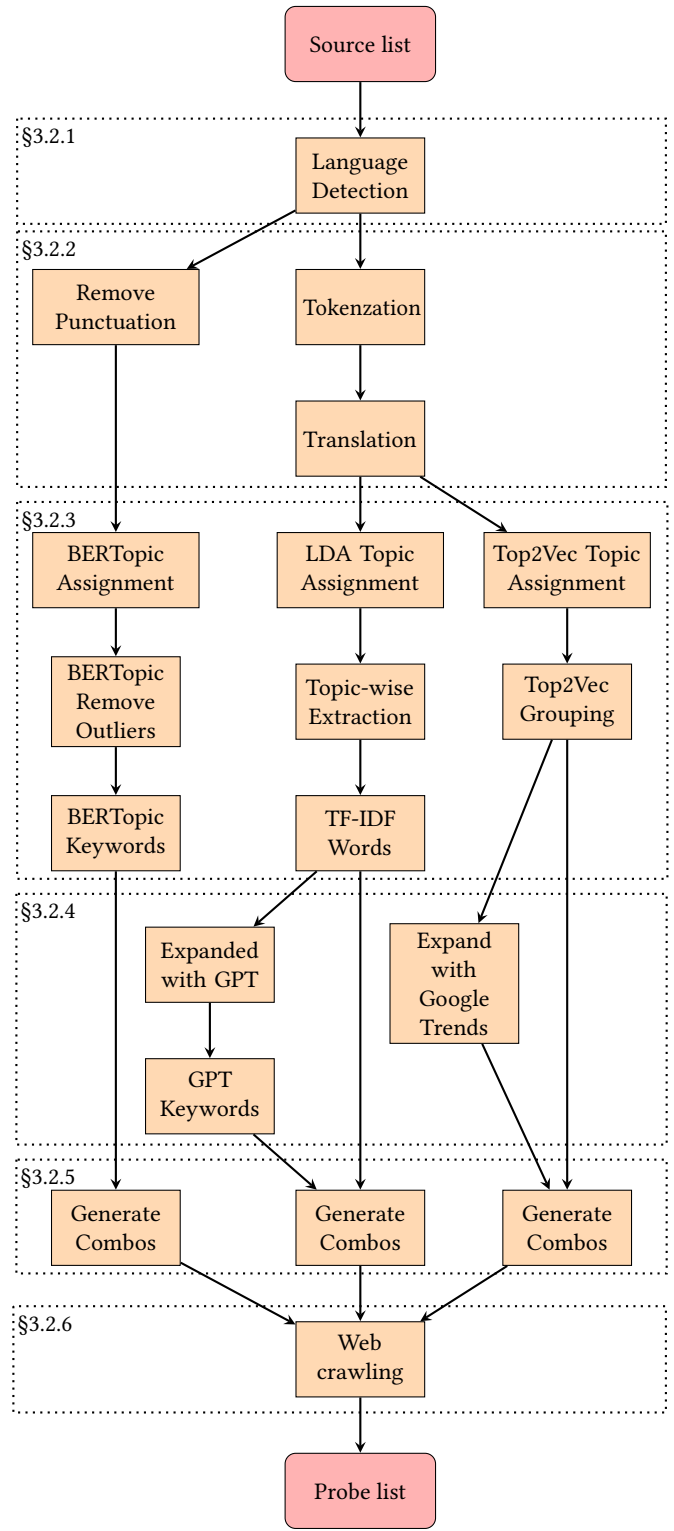
\begin{figure}[thbp]
    \centering
\begin{tikzpicture}[node distance=1.75cm and 2cm]
\node (start) [startstop] {Source list};
\node (pro1) [process, below of=start] {Language Detection};

\node (nlp1) [process, below of=pro1] {Tokenzation};
\node (nlp2) [process, below of=nlp1] {Translation};

\node (lda1) [process, below of=nlp2, text width=2cm] {LDA Topic Assignment};
\node (lda2) [process, below of=lda1] {Topic-wise Extraction};
\node (lda3) [process, below of=lda2] {TF-IDF Words};
\node (lda4a) [process, below of=lda3, xshift=-2cm] {Expanded with GPT};
\node (lda5a) [process, below of=lda4a] {GPT Keywords};
\node (lda6) [process, below of=lda5a, xshift=2cm] {Generate Combos};

\node (t2v1) [process, right of=lda1, xshift=1cm, text width=2cm] {Top2Vec Topic Assignment};
\node (t2v2) [process, below of=t2v1] {Top2Vec Grouping};
\node (t2v3) [process, below of=t2v2, xshift=-1.5cm, yshift=-2cm] {Expand with Google Trends};
\node (t2v4) [process, right of=lda6, xshift=1cm] {Generate Combos}; 

\node (ber1) [process, below of=pro1, xshift=-3.25cm, text width=2cm] {Remove Punctuation};
\node (ber2) [process, left of= lda1, xshift=-1.5cm, text width=2cm] {BERTopic Assignment};
\node (ber3) [process, below of=ber2] {BERTopic Remove Outliers};
\node (ber4) [process, below of=ber3] {BERTopic Keywords};
\node (ber5) [process, left of=lda6, xshift=-1.5cm] {Generate Combos};

\node (crawl) [process, below of=lda6] {Web crawling};
\node (end) [startstop, below of = crawl] {Probe list}; 

\draw [arrow] (start) -- (pro1);
\draw [arrow] (pro1) -- (ber1);

\draw [arrow] (ber1) -- (ber2);
\draw [arrow] (ber2) -- (ber3);
\draw [arrow] (ber3) -- (ber4); 
\draw [arrow] (ber4) -- (ber5); 

\draw [arrow] (pro1) -- (nlp1); 
\draw [arrow] (nlp1) -- (nlp2);

\draw [arrow] (nlp2) -- (lda1);
\draw [arrow] (lda1) -- (lda2);
\draw [arrow] (lda2) -- (lda3);
\draw [arrow] (lda3) -- (lda4a);
\draw [arrow] (lda4a) -- (lda5a);
\draw [arrow] (lda5a) -- (lda6);
\draw [arrow] (lda3) -- (lda6);

\draw [arrow] (nlp2) -- (t2v1);
\draw [arrow] (t2v1) -- (t2v2); 
\draw [arrow] (t2v2) -- (t2v3); 
\draw [arrow] (t2v2) -- (t2v4); 
\draw [arrow] (t2v3) -- (t2v4);

\draw [arrow] (ber5) -- (crawl);
\draw [arrow] (lda6) -- (crawl); 
\draw [arrow] (t2v4) -- (crawl);

\draw [arrow] (crawl) -- (end);

\draw[thick,dotted] ($(ber1.north west)+(-0.2,2)$)  node [left, below, xshift=12] (TextNode1) {\S\ref{subsubsec:language}} rectangle ($(t2v1.north east)+(0.2,4.1)$) ;
\draw[thick,dotted] ($(ber1.north west)+(-0.2,0.5)$)  node [right, below, xshift=12] (TextNode2) {\S\ref{subsubsec:textprocessing}} rectangle ($(t2v1.north east)+(0.2,0.6)$);
\draw[thick,dotted] ($(ber2.north west)+(-0.2,0.4)$)  node [right, below, xshift=12] (TextNode3) {\S\ref{subsubsec:topicassignment}} rectangle ($(t2v2.south east)+(0.47,-1.85)$);
\draw[thick,dotted] ($(ber4.south west)+(-0.46,-0.3)$)  node [right, below, xshift=12] (TextNode4) {\S\ref{subsubsec:topicexpansion}} rectangle ($(t2v4.north east)+(0.46,0.6)$);
\draw[thick,dotted] ($(ber5.north west)+(-0.46,0.4)$)  node [right, below, xshift=12] (TextNode5) {\S\ref{subsubsec:keywordgrouping}} rectangle ($(t2v4.south east)+(0.46,-0.1)$);
\draw[thick,dotted] ($(crawl.north west)+(-3.75,0.2)$)  node [right, below, xshift=12] (TextNode6) {\S\ref{subsubsec:webcrawling}} rectangle ($(crawl.south east)+(3.2,-0.2)$);

\end{tikzpicture}
    \caption{
      \label{fig:leo-thesis}
      Flowchart of the URL generation process. The dashed boxes correspond to different subsections in the body of the text.}
\end{figure}

Figure~\ref{fig:leo-thesis} provides a complete overview of how the
probe list is generated. We start from pages in the sanitized source list,
i.e., the 51,313~URLs that we know are reachable and with actual---
presumably meaningful---content. We then perform the following steps,
on the text content on the page.

\subsubsection{Language detection}
\label{subsubsec:language}

Many pages we examine are not in English. To detect the language of the
page content, which is necessary for topic detection and for translation,
we first use Lingua~\cite{lingua} in high-accuracy mode. For languages
Lingua does not support (15.22\% of the total number of pages), 
we rely on Google's CLD3~\cite{cld3} as a fallback. Eventually, we manage
to identify the language of 99.29\% of the pages. Our corpus spans 103
different languages.

\subsubsection{Text processing} 
\label{subsubsec:textprocessing}
We then prepare the text to perform topic assignment. As we will
discuss later, we will be using three different
techniques for topic assignment in parallel---BERTopic \cite{2022bertopic},
Latent Dirichlet Allocation (LDA,~\cite{blei2003latent}), and
Top2Vec~\cite{top2vec-paper}, which require slightly different
preparations.

BERTopic works with multilingual input and does not need translation,
but requires language identification as part of its parameters. BERTopic
supports 55~languages; page content in unsupported languages is not
passed to that part of the pipeline---we instead only rely on the other
topic assignment techniques for these pages. BERTopic performs best
with sentences rather than individual words, so we do not use standard
tokenization techniques to, e.g., remove stop words. We remove
punctuation, emojis (using the \texttt{demoji} library~\cite{demoji}), leading and
trailing white space, as well as Unicode characters that represent
symbols.

LDA and Top2Vec require similar preparation: tokenization and translation.
Tokenization removes punctuation, stop words, and breaks text into smaller
``core forms'' or tokens. Utilizing open-source libraries and corpora that are
widely used in the linguistics community~\cite{nltk-stopwords,
stopwords-iso, african-stopwords, extra-stopwords, pashto-stopwords,
spacy-lang}, we could identify and remove stop words for 58 languages.
Furthermore, we also use other libraries~\cite{corenlp, amharicprocessor,
hebrew-tokenizer, inltk, classla, rakutenma-python, nlpo3, khmer-nltk} to
tokenize data for languages that do not use spaces or punctuation to
separate between words, such as Chinese, Arabic, or Japanese. 
We then translate the tokens into English. We first run Lingua to filter
out tokens that are already in English (e.g., loan words), and translate
the rest using the Google Translate API~\cite{google-translate-python}. From
this, we get a bag-of-words representation of the web page content.

\subsubsection{Topic Assignment and Keyword Extraction}
\label{subsubsec:topicassignment}
We next map the data representations of the page to a topic that
best describes the content of the page, and extract salient keywords
that characterize this topic. As noted above, we use three different
methods in parallel for this task; this allows us to ultimately generate
as many candidate pages as possible.

\noindent\textbf{BERTopic.} BERTopic produces a topic and a set of keywords
associated with it. We use the cTFIDF model~\cite{bertopic-ctfidf}, which
is similar to the traditional term-frequency/inverse-document-frequency
(TF-IDF) model, but operates at a topic/cluster level instead of the
document level~\cite{bertopic-ctfidf}. Through manual tuning, we discover that the best parameters are
setting the number of words for each topic to 30, and the minimum size
of a cluster to 20 documents. This yields 257 topic
clusters, with an average of 194 documents per group.

\noindent\textbf{Latent Dirichlet Allocation.} Building upon the pre-trained
LDA model by Weinberg et al.~\cite{weinberg2017topics}, we stem the tokens
(i.e., reduce them to their root) with the Porter stemmer from
NLTK~\cite{nltk-stemming} and discard documents with fewer than four words.
This yields 53 potential topics (out of the 64 topics originally identified
by Weinberg et al.~\cite{weinberg2017topics}). We then extract keywords from the documents,
using the TF-IDF library~\cite{scikit-learn-tfidfvectorizer}. This produces
a list of keywords per topic, and their associated TF-IDF scores.

\noindent\textbf{Top2Vec.} Top2Vec~\cite{top2vec-paper} is a third topic
assignment mechanism we use to produce keywords complementary to those
obtained with LDA and BERTopic. Top2Vec produces 232 distinct topics, with
an average of 210 documents per group. Top2Vec does not require additional
keyword extraction through TF-IDF, but produces both a topic assignment and
a list of relevant keywords instead. While BERTopic uses different vector
spaces for topic assignments and keywords, Top2Vec uses a single vector
space~\cite{bertopic-top2vec-comparison}.

\subsubsection{Topic Expansion}
\label{subsubsec:topicexpansion}

Our method for generating new URLs is similar to snowball
sampling---starting from a seed and expanding from it. Snowball
sampling, however, might lead to biases. In particular, while it is
effective at producing new candidate pages on topics 
observed before, snowball sampling is not suited to discovering new
topics; this is particularly problematic for us, since censorship
evolves over time, frequently in response to shifting dynamics in news
events. We mitigate this issue in two ways.

\noindent\textbf{Asking ChatGPT for suggestions.} Large language models,
such as ChatGPT, use large text corpora to attempt
to answer questions from users. As such, they are 
well suited to suggest related topics from an existing corpus.
We expand topics found by our LDA algorithm using ChatGPT,
specifically, the \texttt{gpt-3.5-turbo} version. We describe in
Appendix~\ref{appendix:gpt-prompt} the prompt we use, to ask ChatGPT how
to expand the set of topics coming from LDA analysis.

\noindent\textbf{Using Google Trends as a complementary source.} 
The source list relies on fairly aged inputs---some test lists dates
back to 2007. To ensure that our probe list have some current inputs,
we complement the topics found with Top2Vec with related input from Google
Trends~\cite{google-trends}.
Specifically, for each topic produced by Top2Vec, we extract the two
most relevant keywords (using cosine similarity to the topic) and feed
them into Google Trends to obtain related keywords over the preceding
five years before May 2023. Google Trends responds with ``Top'' trends---i.e., those
that have consistently ranked high, and ``rising'' trends---i.e.,
trends that are becoming more popular at the time the query is made. 
We limit our search to a maximum of 40 new keywords per topic; we get 36 on
average.

\subsubsection{Generating Search Strings: Keyword Grouping}
\label{subsubsec:keywordgrouping}
The next step, common to the whole pipeline, is to generate combinations
of keywords that will eventually be fed into a search engine to discover
new pages of interest. We manually experimented and discovered that using four to nine keywords was optimal. Four or fewer keywords led to very generic, mostly irrelevant results; nine or more keywords was too specific and yielded no result. Furthermore, keyword order in
the search matters, returning different results, and so we try various different permutations.

We also discovered that ``tiering'' improves search results.
Specifically, we divide the keywords associated with each topic into
four tiers, in decreasing order of semantic affinity to the topic. For
the desired number of keywords (between four and nine, as discussed
above), we then draw from each tier probabilistically: we select tier-1
keywords with a probability of 0.25 to 0.5; tier~2 with a probability of
0.05 to 0.4; tier~3 with a probability of 0.05 to 0.2; the remaining is
filled with tier-4 keywords if necessary.
The idea is to favor slightly more representative keywords,
which ensures diverse inputs. To that effect, we also limit ourselves
to at most three tier-1 keywords. Thus, higher tier (more relevant) keywords
appear more frequently in the search combinations, while maintaining a random factor
using keywords from other tiers.

\subsubsection{Google Search Web Crawling}
\label{subsubsec:webcrawling}
The keyword grouping process results in 14,450 query combinations.
We
feed those into the Google Search API~\cite{google-custom-search}
to obtain URLs to include in our probe list related to these topics. For
each keyword combination, the API returns a default of 10~results.
When results cannot be found or appear rare, Google sometimes returns a
``spell-corrected'' query, which we recursively call once to maximize
the number of URLs generated. When a search string yields no result and
no spell-corrected query, we reduce the keyword combination size by 20\%
and retry until we obtain search results. The process generates a total
of 160,981~URLs, which after removing duplicates, reduces to a probe list
containing 119,272~URLs.

\section{Testing the Probe List}
\label{sec:testing_probe_list}

Using this new probe list of 119,272 candidate URLs, we attempt to access
each URL from thirteen vantage points in eleven cities, as
detailed in Table~\ref{tab:locations}. Our selection spans several
locations with varying degrees of Internet freedom, as categorized by
Freedom House~\cite{FreedomStatus}. This includes vantage points in locations
classified as ``free'' (Japan, Taiwan, United Kingdom, France, USA),
``partly free'' (Hong Kong, India, Singapore), and ``not free'' (P.R.
China).

In Asia, we focus on diverse environments, ranging from locations with
known information controls (Beijing, Shanghai) to those typically
associated with greater Internet freedom (Tokyo, Taipei). This contrast is
crucial for our analysis, as it allows us to observe potential censorship
patterns within the same continent. Similarly, by including locations in
countries such as the United Kingdom and France, both classified as
``free,'' we aim to build a more complete baseline control against which to evaluate our probe list.

We clean the URLs generated from our set of 119,272 URLs by removing
any text in the URL after a comma (including the comma). We also escape
single quotation marks. After cleaning up, we have 17 repeated URLs, so
this results in 119,255 unique URLs.\footnote{Also, 14 URLs contain the
\$ character. We noticed after the fact that these interact poorly
with our test environment due to shell variable expansion---they 
consistently returned 403 Forbidden errors. This glitch affects less
than 0.0001\% of our dataset.}

\begin{table}[htbp]
  \centering
  \caption{
    \label{tab:locations}
    Testing vantage points and their freedom status evaluated by the Freedom House~\cite{FreedomStatus}.}
  \begin{adjustbox}{width=\columnwidth, center}  
  \begin{tabular}{lll}
    \toprule
    \textbf{Country/Region} & \textbf{Freedom Status} & \textbf{Test Location} \\
    \midrule
    China & Not Free & Beijing \\
          &          & Shanghai \\
    Hong Kong & Partly Free & Hong Kong 1 \\
              &             & Hong Kong 2 \\
    India & Partly Free & Mumbai \\
    Japan & Free & Tokyo \\
    Singapore & Partly Free & Singapore \\
    Taiwan & Free & Taipei \\
    \midrule
    United Kingdom & Free & London \\
    France & Free & Paris \\
    \midrule
    US-East  & Free & Academic Network \\
    US-West 1& Free & Commercial Network\\
    US-West 2& Free & Commercial Network\\
    \bottomrule
  \end{tabular}
  \end{adjustbox}
\end{table}

\begin{table}[htbp]
  \centering
  \caption{
    \label{tab:urlDomainCounts}
   Number of unique URLs and domains in each dataset.}
  \begin{adjustbox}{width=\columnwidth, center}
  \begin{tabular}{lrr}
    \toprule
    \textbf{Dataset} & URLs & Domains\\
    \midrule
    \textbf{Source List} & 139,957 & 106,878\\
    \textbf{Probe List} & 119,255 & 35,147\\
    \textbf{Probe List (New Domains Only)} & 71,960 & 32,543\\
    \textbf{Probe List (Domains in Source List Only)} & 47,295 & 2,604\\
    \bottomrule
  \end{tabular}
  \end{adjustbox}
\end{table}

To help establish a baseline representing an uncensored environment, our US-East measurements are
conducted from the Carnegie Mellon University network in the United States, where
we do not expect any censorship to take place. US-West measurements originate from commercial servers
of two different hosting companies in Silicon Valley. For testing in other
regions, we utilized Virtual Private Servers (VPS) under our direct
control, strategically located in different cities worldwide, as
outlined in Table~\ref{tab:locations}.\footnote{US-West 1 and US-West 2 are located in the same city.} By routing our test traffic through
the networks of these diverse locations, we could subject our URL requests
to any regional restrictions or policies enforced by corresponding
local authorities.\footnote{Singapore may be a special
case, which is addressed in~\sectionref{sec:discussion}.}

For each tested URL, we collected the web page response details,
specifically the HTTP status code and \texttt{curl} exit code. Our goal
is to assess if the page could be accessed from the various locations, and
if not, to identify the nature of the error or connection failure
encountered. In some cases, web pages may appear inaccessible due to
server-side blocking mechanisms specifically refusing \texttt{curl}
requests. To remedy this, we supplemented our measurements with data
gathered using the OONI Probe~\cite{ooni-cli} to gain deeper insights into
potential blocking, providing more comprehensive information about
censorship and network interference (\sectionref{sec:ooni-tests}).

\subsection{Response Types}

Attempting to access any of the URLs results in three broad categories
of responses:

\begin{enumerate}
\item \textbf{Accessible.} These are URLs that return a 2XX HTTP status
code, or a 3XX status code. We do not click on any links or follow any
redirects, so we identify redirects as accessible for
the sake of this experiment.\footnote{Certain ISPs block pages using 302
Redirects \cite{ververis2015understanding}. As such, our results are a
conservative lower bound of actual censorship.}

\item \textbf{Inaccessible.} These are URLs that return a non-zero \texttt{curl}
exit code, such as would result from a DNS error, port not connectable,
invalid certificate, or formatting error. For results with an exit
code of 28 (timeout), we specified a limit of 30~seconds. 15~seconds
is the amount of time that Cloudflare chose as its timeout limit to
establish a connection before resulting in Error 522: connection timed out~\cite{CF-5xx}, 
so we double this number for an upper bound. Based on prior
work \cite{NielsenHowLong}, we believe this number is a reasonable
estimate for the maximum amount of time a human user may want to wait
for a page to load.

\item \textbf{Error.} These URLs return a status code that is
not in the 200s or 300s. The majority of status codes in this category are
in the 4XX range. These generally imply client-side issues or server-side blocking such as Unauthorized
(403) or Not Found (404).
\end{enumerate}

\begin{figure}[htbp]
\centering
\includegraphics[width=\columnwidth]{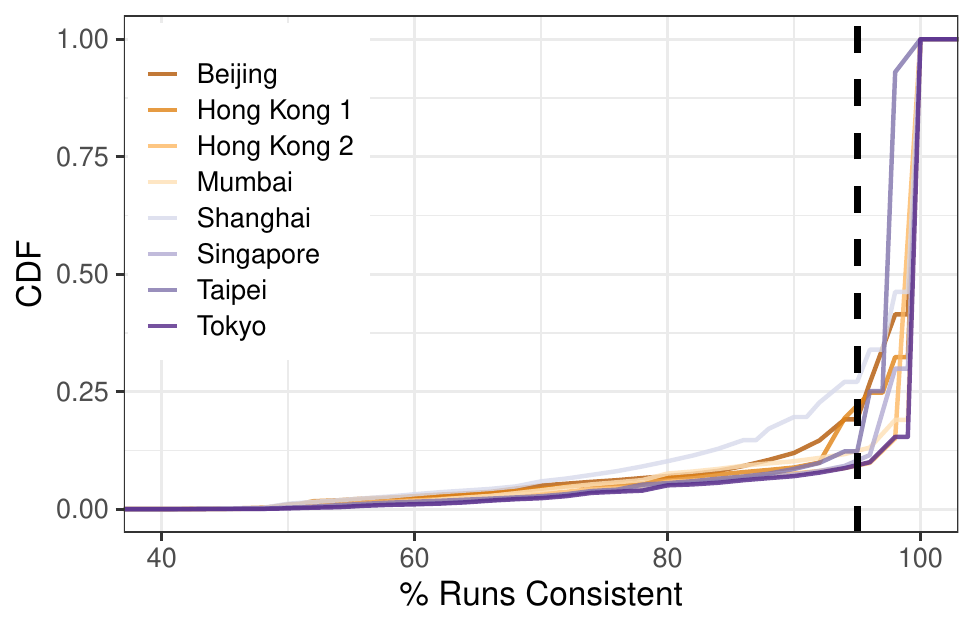}
  \caption{The frequency (CDF) of percentage of consistent runs of URLs for each location outside the baseline. The vertical line (at 95\%) is the cutoff for considering responses from a URL to be inconsistent and excluding it from our data.}
\label{fig:consistency}
\end{figure}

To account for uncontrollable factors such as transient errors or packet
loss, we repeat our measurements 50 times in each vantage point (apart
from US-West 1 and US-West 2, where measurements were only repeated 8 times)
between November 2023 and March 2024. In Figure~\ref{fig:consistency},
we plot the cumulative distribution of the number of runs that produce
consistent results, for each location. For most locations, the c.d.f.
presents a ``bend'' around 95\%. Thus,
we consider the results from a URL to be consistent if accessing the URL returns
the same code over 95\% of the time, and classify the result from that URL
as the code that it consistently returns. This allows
us to distinguish between spurious errors (potentially indicative of
transient problems) and consistent errors (potentially indicative of
censorship).

Unless otherwise noted, our analysis throughout this paper will use the
results from only those URLs that are consistent.

\subsection{Differences with Baseline (``Diff'')}

We make no claims about the content of the pages, but aim to
evaluate whether pages generated by our automated pipeline are
potentially blocked in various locations. Therefore, we want to identify
pages that are inaccessible only in certain locations, and not pages
that are inaccessible because they no longer exist, are unavailable,
or are inaccessible everywhere. To do so, we do not evaluate pages whose responses overlap with the
``baseline.''

Our ``baseline''---an average view across locations with high freedom scores, where pages are less likely to be censored---
is determined from an aggregate of the five vantage points in the US, UK, and France. If all vantage points agree on the result (``accessible'', ``inaccessible'', ``error'') of an URL, then we classify that page as such in the baseline set.
If our baseline vantage points do not all agree on the result of a URL, then it is not included in the baseline set.

We determine differences in each location by examining which URLs return results (accessible,
inaccessible, or result in an error) that differ from their result in the baseline. 
If \texttt{example.com} is accessible
in Singapore but results in an error in our baseline, then
\texttt{example.com} is different from our baseline and in futher analysis evaluated as an accessible site in Singapore. However, if
\texttt{example.com} returns status code 4XX in Singapore and status
code 5XX in our baseline set, then it is 
not considered to be different, as it results in an error in both
locations. It thus would not be further evaluated in our Singapore dataset. We use general categories rather than specific codes to
determine differences, as the particular type of error or reason for being
inaccessible varies, but, to a general user trying to access
these pages, it remains inaccessible. In short, unless otherwise stated, we only consider pages
that give different results when accessed in these locations compared to
the baseline in our analysis, resulting in a lower bound estimate of differences.

If the URL is not in the baseline set (i.e. our five baseline vantage points do not agree on its result), it is considered to differ from the baseline if it consistently returns some result in a vantage point.

\section{Probe List Evaluation}
\label{sec:probe_list_eval}

We next evaluate the utility of our generated probe list across the testing
vantage points, highlighting the effectiveness of our system in identifying
previously unknown domains outside the source list that are potentially
censored. As the majority of tested URLs are prefixed with
\texttt{https://}, we primarily focus our analysis on the
domains of the tested URLs rather than their paths.

\subsection{Overall Result}

\begin{table*}[htbp]
  \centering
  \caption{
    \label{tab:countsFullAD}
    Within each vantage point, the number of URLs and domains with consistent results for each response type, with associated proportions in parentheses. As there may be multiple URLs from one domain, a domain may return multiple different responses depending on the URL, and thus be counted in more than one category. Thus, the sum of percentages may exceed 100\%.}
  \begin{adjustbox}{width=\textwidth}
        \begin{tabular}{lrrrlrrrr}
        \toprule
        \textbf{Location} &  URL (Full) &  URL (Full) &  URL (Full) &  URL (Full) &  Dom (Full) &  Dom (Full) &  Dom (Full) & Dom (Full)  \\
        All Domains & Accessible & Inaccessible & Error & Total & Accessible & Inaccessible & Error & Total\\
\midrule
\textbf{London}         & 95,017 (87.36\%) & 2,316 (2.13\%) & 11,428 (10.51\%) & 108,761 & 30,042 (91.60\%) & 769 (2.34\%) & 2,836 (8.65\%)& 32,796 \\
\textbf{Paris}          & 94,945 (87.40\%) & 1,965 (1.81\%) & 11,724 (10.79\%) & 108,634 & 29,851 (91.16\%) & 775 (2.37\%) & 2,975 (9.08\%)& 32,747 \\
\textbf{US-East}        & 96,813 (90.22\%) & 1,027 (0.96\%) & 9,472 (8.83\%)   & 107,312 & 30,768 (93.83\%) & 527 (1.61\%) & 2,291 (6.99\%)& 32,791 \\
\textbf{US-West 1}      & 98,439 (86.13\%) & 2,095 (1.83\%) & 13,757 (12.04\%) & 114,291 & 31,020 (91.33\%) & 857 (2.52\%) & 3,068 (9.03\%)& 33,966 \\
\textbf{US-West 2}      & 97,102 (85.25\%) & 2,468 (2.17\%) & 14,338 (12.59\%) & 113,908 & 30,942 (91.26\%) & 888 (2.62\%) & 3,053 (9.00\%)& 33,905 \\
\hline
\rowcolor{lightgray}\textbf{Beijing}        & 64,518 (66.88\%) & 19,537(20.25\%) & 12,414 (12.87\%) & 96,469 & 23,402 (80.66\%) & 2,453 (8.45\%) & 3,929 (13.54\%)& 29,014 \\
\textbf{Hong Kong 1}    & 84,097 (87.45\%) & 1,849 (1.92\%) & 10,222 (10.63\%) & 96,168 & 28,109 (91.28\%) & 765 (2.48\%) & 2,647 (8.60\%)& 30,793 \\
\textbf{Hong Kong 2}    & 93,047 (85.51\%) & 3,319 (3.05\%) & 12,445 (11.44\%) & 108,811 & 29,329 (90.05\%) & 1,102 (3.38\%) & 2,997 (9.20\%)& 32,571 \\
\textbf{Mumbai}         & 91,549 (86.97\%) & 2,083 (1.98\%) & 11,636 (11.05\%) & 105,268 & 29,623 (91.05\%) & 839 (2.58\%) & 2,916 (8.96\%)& 32,536 \\
\rowcolor{lightgray}\textbf{Shanghai}       & 55,969 (64.40\%) & 19,447(22.37\%) & 11,498 (13.23\%) & 86,914 & 22,398 (80.74\%) & 2,517 (9.07\%) & 3,552 (12.80\%)& 27,740 \\
\textbf{Singapore}      & 93,800 (86.64\%) & 2,543 (2.35\%) & 11,926 (11.02\%) & 108,269 & 29,585 (90.83\%) & 870 (2.67\%) & 2,982 (9.15\%)& 32,573 \\
\textbf{Taipei}         & 92,228 (88.19\%) & 1,564 (1.50\%) & 10,789 (10.32\%) & 104,581 & 29,034 (91.45\%) & 739 (2.33\%) & 2,787 (8.78\%)& 31,748 \\
\textbf{Tokyo}          & 95,194 (87.47\%) & 2,051 (1.88\%) & 11,590 (10.65\%) & 108,835 & 29,995 (91.57\%) & 803 (2.45\%) & 2,827 (8.63\%)& 32,758 \\
\bottomrule
        \end{tabular}
    \end{adjustbox}
\end{table*}

\begin{table*}[htbp]
  \centering
  \caption{
    \label{tab:pctDiffNDvAD}
    Percent of URLs and domains in each response type that were different from the baseline that came from domains not in the source list. Diff means the number of URLs or domains in each category that were not in that category in the baseline. For example, 78\% (91\%) of the domains that were inaccessible in Beijing (Mumbai) were from domains not in our source list. Overall, 85\% (87\%) of domains that returned different results from the baseline in Beijing (Mumbai) were from new domains not in our source list.}
  \begin{tabular}{lrrrlrrrr}
    \toprule
    \textbf{Location}  & URL (Diff) &  URL (Diff) &  URL (Diff) &  URL (Diff) & Dom (Diff) &  Dom (Diff) &  Dom (Diff) & Dom (Diff)\\
    New/All & Accessible \% & Inaccessible \% & Error \% & Total \% & Accessible \% & Inaccessible \% & Error \% & Total \%\\
\midrule
\rowcolor{lightgray}\textbf{Beijing} & 49.73 & 25.03 & 80.39 & 41.63 & 87.15 & 78.06 & 90.26 & 85.40 \\
\textbf{Hong Kong 1} & 47.84 & 81.05 & 74.70 & 57.36 & 86.79 & 92.37 & 89.58 & 88.66 \\
\textbf{Hong Kong 2} & 47.48 & 75.10 & 71.08 & 59.64 & 86.01 & 91.78 & 88.39 & 88.32 \\
\textbf{Mumbai} & 48.02 & 81.25 & 74.93 & 58.97 & 86.61 & 90.69 & 87.37 & 87.84 \\
\rowcolor{lightgray}\textbf{Shanghai} & 67.38 & 27.51 & 80.30 & 43.34 & 87.69 & 77.45 & 90.66 & 85.02 \\
\textbf{Singapore} & 49.24 & 69.62 & 75.36 & 59.78 & 85.78 & 90.09 & 88.38 & 87.63 \\
\textbf{Taipei} & 51.29 & 93.56 & 74.73 & 58.93 & 85.59 & 93.24 & 90.14 & 88.12 \\
\textbf{Tokyo} & 51.86 & 70.47 & 74.02 & 59.57 & 86.66 & 89.97 & 88.55 & 88.02 \\
\bottomrule
  \end{tabular}
\end{table*}

Our measurements revealed evidence of potential censorship, with varying
rates of URLs and domains being (in)accessible across the tested
locations, as detailed in Table \ref{tab:countsFullAD}. Notably, Beijing
and Shanghai exhibited significantly higher numbers of inaccessible cases
compared to other vantage points. This disparity in connectivity suggests
the presence of regional disparities in web accessibility, potentially
attributable to nation-state censorship practices.

However, some of the inaccessible pages or
pages that returned errors may not necessarily be indicative of intentional
blocking due to content sensitivity. These issues could arise from general
website problems, such as the site no longer existing or being temporarily
unavailable, leading to errors (e.g., HTTP 404) regardless of the access
location.

To remedy this issue and isolate potential instances of censorship, recall that we
establish a baseline by aggregating measurements from five vantage points
with high Internet freedom scores. This baseline served as an effective
reference point, allowing us to identify pages that exhibited different
accessibility patterns compared to the blocking behaviors commonly observed
in censored environments.

Consequently, our subsequent analysis focused exclusively on the pages that
yielded responses different from those observed in the baseline condition.
By contrasting against this baseline, we can more accurately pinpoint
instances where inaccessibility might be attributable to censorship rather
than general website issues.

\subsection{Known Domains: Overlap with Source List} 

While our system aims to generate previously \textit{unknown} instances of
potential censorship (i.e. URLs that are not in the source list), it is
inevitable that some of the generated pages overlap with those in the
source list, since our probe list generation method uses keyword-browsing
combinations to find candidate pages. We call these overlapping cases
``known domains''.

Our system is effective at finding URLs from known domains. Of the 119,255
URLs, 47,295 (39.7\%) originate from domains present in our full 139K URL
source list. These URLs are from 2,604 (7.4\%) of known domains. While known
domains comprise less than 7.5\% of the total domains in our dataset, these
domains contribute to 15\% of the differences between domains in our
measurements in Beijing and Shanghai from our baseline.

The higher proportion of domains being inaccessible in China may stem from
biases in our source list and our system's method of generating candidate
pages. As shown in Table~\ref{tab:PctFullKD} (in
Appendix~\ref{appendix:moar_tables}), which displays the proportion of
pages generated from domains already present in our source list, a much
greater portion (over 20\%) of known domains (including those that
return the same response as in the baseline) are inaccessible from China
compared to most other locations ($<$4.5\%). Indeed, only around 68\% of
known domains are accessible when accessed from Beijing and Shanghai,
whereas this proportion exceeds 90\% elsewhere. These results suggest that
some pages on the source list, despite being outdated, remain blocked,
potentially because their topics continue to be sensitive, making them more
likely to appear in search results from related keyword combinations.
It also suggests that while the original source list appears to accurately identify domains and
URLs blocked in China, it may be biased toward this locality. The majority
($>$70\%) of URLs inaccessible in Beijing and Shanghai came from
known domains, while this proportion was less than 31\% in all other locations.
This implies that for potentially blocked pages, our
generated probe list is heavily influenced by the source list, generating
many URLs from known blocked domains. Nonetheless, given the changes in 
pages online and our methods of augmenting keywords, our system can still
discover new domains that may be subject to censorship.

\subsection{New Domains}

We classify domains as ``new domains'' if they appear in the probe list but
not in the source list. This allows us to evaluate the utility of our
system in the automated generation and detection of previously unknown
censorship instances. To establish a conservative lower bound on the number
of new domains that we discover through our automated process, we focus on
domains that did not appear in the set of 106K unique domains from the full
source list of 139K URLs rather than from the subset of 51K URLs used to
generate our probe list. Remarkably, out of the 35,000 unique domains in
our probe list, over 32,500 were not present in our source list. Hence, we
filter our dataset down to 71,960 unique URLs originating from these
previously unseen domains for all further
analysis~(Table~\ref{tab:urlDomainCounts}). Unless otherwise noted, all
analysis from this point forth is conducted on this subset of ``new
domains.''

\begin{table*}[htbp]
  \centering
  \caption{
    \label{tab:pctDiffND}
    Of all the URLs or domains that gave a response different (Diff) from that given in the baseline (given in the first row), the count and percentage of URLs or domains within each response type by location. For example, of all URLs returned different results from the baseline in Shanghai, 42\% of these URLs differed because they were inaccessible. 36\% of all domains in Shanghai that differed from the baseline were inaccessible. Since different URLs from the same domain may return different types of codes, the sum of percentages may exceed 100\%.}
\begin{adjustbox}{width=\textwidth}
        \begin{tabular}{lrrrlrrrr}
        \toprule
        \textbf{Location} & URL (Diff) &  URL (Diff) &  URL (Diff) &  URL (Diff) & Dom (Diff) &  Dom (Diff) &  Dom (Diff) & Dom (Diff) \\
        New Domains & Accessible & Inaccessible & Error & Total & Accessible & Inaccessible & Error & Total\\
\midrule
\textbf{Baseline}   & 54,369 (90.78\%) & 673 (1.12\%) & 4,851 (8.10\%) & 59,893 & 26,550 (93.83\%) & 411 (1.45\%) & 1,847 (6.53\%) & 28,295 \\
\hline
\rowcolor{lightgray}\textbf{Beijing}        & 3,082 (23.26\%) & 4,717 (35.60\%) & 5,451 (41.14\%) & 13,250 & 1,289 (26.62\%) & 1,619 (33.43\%) & 2,011 (41.52\%) & 4,843 \\
\textbf{Hong Kong 1}    & 3,317 (56.05\%) & 924 (15.61\%) & 1,677 (28.34\%) & 5,918 & 1,643 (62.71\%) & 339 (12.94\%) & 688 (26.26\%) & 2,620 \\
\textbf{Hong Kong 2}    & 3,265 (41.15\%) & 1,873 (23.60\%) & 2,797 (35.25\%) & 7,935 & 1,654 (53.48\%) & 603 (19.50\%) & 891 (28.81\%) & 3,093 \\
\textbf{Mumbai}         & 3,489 (50.41\%) & 1,053 (15.21\%) & 2,379 (34.37\%) & 6,921 & 1,720 (60.41\%) & 370 (13.00\%) & 816 (28.66\%) & 2,847 \\
\rowcolor{lightgray}\textbf{Shanghai}       & 2,415 (19.96\%) & 5,136 (42.44\%) & 4,550 (37.60\%) & 12,101 & 1,211 (26.94\%) & 1,635 (36.37\%) & 1,719 (38.23\%) & 4,496 \\
\textbf{Singapore}      & 3,249 (46.49\%) & 1,196 (17.11\%) & 2,544 (36.40\%) & 6,989 & 1,581 (57.20\%) & 382 (13.82\%) & 867 (31.37\%) & 2,764 \\
\textbf{Taipei}         & 4,313 (63.27\%) & 712 (10.44\%) & 1,792 (26.29\%) & 6,817 & 1,799 (65.56\%) & 276 (10.06\%) & 731 (26.64\%) & 2,744 \\
\textbf{Tokyo}          & 3,855 (55.31\%) & 864 (12.40\%) & 2,251 (32.30\%) & 6,970 & 1,774 (64.37\%) & 323 (11.72\%) & 727 (26.38\%) & 2,756 \\
\bottomrule
        \end{tabular}
    \end{adjustbox}
\end{table*}

In both Beijing and Shanghai, a higher portion ($>$80\%) of inaccessible new domains were accessible (or
returned an error) in the baseline, while in other vantage points this was less than 50\%, except for Hong Kong 2 at
60\% (Table~\ref{tab:pctDiffvFullND}). This implies that we are observing a
higher rate of potential blocks from candidate pages in our probe list from
locations in China. This suggests that our system is good at finding
potentially new blocked pages for China, although the results are less
conclusive for other locations.

\begin{table*}[htbp]
  \centering
  \caption{
    \label{tab:pctDiffvFullND}
    Percentage of URLs and domains with responses different from baseline (Delta), only looking at new domains. So 29\% of new domains that returned an error in Tokyo did not return an error in our baseline. However, only 9\% of new domains tested in Tokyo returned a response that differed from our baseline's response.}
  \begin{tabular}{lrrrlrrrr}
    \toprule
    \textbf{Location} &  URL &  URL &  URL &  URL & Domain &  Domain &  Domain & Domain \\
    Delta & Accessible \% & Inaccessible \% & Error \% & Total \% & Accessible \% & Inaccessible \% & Error \% &Total \%\\
\midrule
\rowcolor{lightgray}\textbf{Beijing} & 6.84 & 88.93 & 57.19 & 22.13 & 5.89 & 82.43 & 56.86 & 18.08 \\
\textbf{Hong Kong 1} & 6.38 & 61.52 & 26.64 & 9.90 & 6.32 & 49.06 & 29.34 & 9.19 \\
\textbf{Hong Kong 2} & 5.83 & 73.65 & 36.77 & 11.99 & 6.10 & 60.36 & 33.78 & 10.27 \\
\textbf{Mumbai} & 6.14 & 61.94 & 33.17 & 10.54 & 6.26 & 49.07 & 31.79 & 9.44 \\
\rowcolor{lightgray}\textbf{Shanghai} & 5.87 & 89.10 & 51.30 & 21.71 & 5.78 & 81.75 & 53.62 & 17.55 \\
\textbf{Singapore} & 5.72 & 64.09 & 34.49 & 10.58 & 5.78 & 49.10 & 32.95 & 9.17 \\
\textbf{Taipei} & 7.78 & 51.93 & 27.47 & 10.77 & 6.71 & 41.32 & 29.61 & 9.35 \\
\textbf{Tokyo} & 6.66 & 56.32 & 31.78 & 10.48 & 6.39 & 45.05 & 29.14 & 9.09 \\
\bottomrule
  \end{tabular}
\end{table*}

\subsection{OONI Probe Tests}
\label{sec:ooni-tests}

To augment our analysis, we conducted additional tests using OONI
Probe~\cite{ooni-cli} at each vantage point. This allowed us to gain deeper
insights and more detailed measurements in comparison to the results
obtained from \texttt{curl}. For each location, we specifically tested the
list of URLs associated with new domains that have different
results from our baseline measurements.

\noindent\textbf{Agreement with \texttt{curl} results.} 
The results obtained from the OONI Probe measurements generally aligned
with our \texttt{curl}-based testing. We considered the results to be in
agreement if \emph{1)} both \texttt{curl} and OONI concurred that a URL was
accessible, \emph{or} if  \emph{2)} \texttt{curl} failed to connect (returning an
inaccessible or error code) while OONI detected an anomaly (DNS,
TCP/IP, or HTTP). The distribution of measurements exhibiting this
agreement can be found in Table~\ref{tab:ooniCounts}. Through this combined
analysis, we identified 1,490 unique domains that potentially faced
blocking, as they remained inaccessible for over four months of
\texttt{curl} measurements and triggered anomalies in the OONI tests.

A significant portion ($>$70\%) of the domains that our \texttt{curl} tests marked as
inaccessible in Beijing or Shanghai also triggered anomalous results indicative of potential
blocking in the OONI measurements (Table~\ref{tab:ooniPercents} in Appendix~\ref{appendix:moar_tables}). Notably, in Beijing and Shanghai, over
1,200 domains not present in our original source list returned anomalies
detected by OONI and consistently failed to connect via \texttt{curl}. Among
these, 1,068 unique domains exhibited anomalous OONI results in both
locations, with a total of 1,355 domains affected in at least one of the
two cities. This finding strongly suggests that our discovery method for
compiling the probe list successfully identified previously unknown domains
that may be subject to blocking in these regions.

Moreover, we observed overlaps with other known blocked
pages~\cite{GFWatch_dashboard} in these locations but were not part of our initial source list, 
such as \texttt{genius.com} and \texttt{huggingface.co}.
We also uncovered domains that, while not on known blocked
lists, thematically align with potential censorship targets, such as
\texttt{governmentjobs.com} and \texttt{rilot.com} (Rhode Island Lottery),
as well as numerous pornography and adult sites.

\noindent\textbf{Where do the newly discovered pages come from?} The
majority of inaccessible URLs that also triggered anomalies in OONI were
generated by Top2Vec-Trends (approximately 58\%), followed by LDA-TFIDF and
Top2Vec (each accounting for 13-14\% of the total number of URLs). This
observation suggests that utilizing Top2Vec-Trends is an effective approach
for identifying potentially blocked pages and updating probe lists.

\subsection{Comparison with Previous Probe List Generation Efforts}

Our approach demonstrates
over 10 times higher efficacy in discovering potentially blocked domains
compared to similar prior efforts. While
FilteredWeb~\cite{Darer2017FilteredWebAF} discovered 4.11 blocked
domains per 1,000 domains crawled, our system identified 45.79 potentially
blocked domains per 1,000 domains crawled. Moreover, compared to the work
by Hounsel et al.~\cite{autoblocklist}, which found 1,255 blocked domains
in crawls of 1,000,000 URLs, our approach uncovered 1,490 potentially
blocked domains in crawls of just 71,960 URLs. Remarkably, the vast
majority (1,473) of these 1,490 newly discovered domains are not part of
the 1,255 domains found in~\cite{autoblocklist}, suggesting the efficacy of
our system in identifying domains and content that differ from previous efforts.

\subsection{Verification Against the GFW} 

To further validate our
findings, we examined the potentially blocked domains identified in Beijing
and Shanghai by testing them directly against the GFW's DNS, HTTP, and
HTTPS filters, with each domain tested at least three times. The vast
majority (over 90\%) of domains exhibiting DNS anomalies as identified by OONI were
indeed blocked by the GFW's DNS filter~\cite{Anonymous2020:TripletCensors,
USESEC21:GFWatch} (429/457 for Beijing and 422/461 for Shanghai). A smaller number of domains where OONI has detected TCP/IP (15/657
for Beijing, 14/669 for Shanghai) or HTTP-Failure anomalies (12/90 for
Beijing, 6/91 for Shanghai) can also be confirmed to be blocked by the GFW,
based on known blocking patterns~\cite{Wang2017YourSI, Bock2021EvenCH}. In total, 527 unique
domains between Beijing and Shanghai were detected to be blocked by
the GFW.

While the majority of domains exhibiting TCP/IP anomalies detected by OONI
were not directly present on the GFW's blocklists, our further
investigation suggests that the GFW also blocks the hosting servers (IP
addresses) of these domains, rather than the domains themselves.
Consequently, these domains remained inaccessible due to the blocking of
their hosting IP addresses.

To that end, given the agreement between our \texttt{curl} tests, OONI
measurements, and the GFW tests, we are confident that we have identified
over 500 domains that are almost certainly censored in China, with an
additional 718 domains suspected of being blocked based on the observed
TCP/IP anomalies. This observation also underscores the significance of
collateral damage caused by IP-based blocking.

\begin{table*}[htbp]
  \centering
  \caption{
    \label{tab:ooniCounts}
    Number of URLs and domains with each type of result from tests with the OONI Probe that triggered anomalies. The URLs tested on OONI were those that consistently returned the same inaccessible or error response with \texttt{curl} over a period of 4 months, to reduce the chance of transient network issues causing false positives. There are four types of anomalies that may be indicative of potential censorship (DNS, TCP/IP, HTTP-Failure, and HTTP-Diff). The percentages in these columns indicate the proportion of anomalous measurements each type makes up. Since different URLs from the same domain may return different types of codes, the sum of percentages may exceed 100\%.}
    \begin{adjustbox}{width=\textwidth}
\begin{tabular}{lrrrrlrrrrr}
\toprule
 Location & URL & URL & URL & URL & URL & Domain & Domain & Domain & Domain & Domain\\
 & DNS & TCP/IP & HTTP-Failure & HTTP-Diff & Total & DNS & TCP/IP & HTTP-Failure & HTTP-Diff & Total \\
\midrule
\rowcolor{lightgray}\textbf{Beijing}        & 1,403 (45.64\%) & 1,496 (48.67\%) & 155 (5.04\%) & 20 (0.65\%)& 3,074 & 457 (38.08\%) & 657 (54.75\%) & 90 (7.50\%) & 11 (0.92\%)& 1,200 \\
\textbf{Hong Kong 1}    & 2 (0.38\%) & 506 (96.20\%) & 18 (3.42\%) & 0 (0.00\%)& 526 & 1 (0.44\%) & 213 (94.67\%) & 11 (4.89\%) & 0 (0.00\%)& 225 \\
\textbf{Hong Kong 2}    & 1 (0.18\%) & 521 (94.38\%) & 28 (5.07\%) & 2 (0.36\%)& 552 & 1 (0.31\%) & 299 (94.03\%) & 16 (5.03\%) & 2 (0.63\%)& 318 \\
\textbf{Mumbai}         & 2 (0.51\%) & 361 (92.56\%) & 26 (6.67\%) & 1 (0.26\%)& 390 & 2 (1.11\%) & 163 (90.56\%) & 15 (8.33\%) & 1 (0.56\%)& 180 \\
\rowcolor{lightgray}\textbf{Shanghai}       & 1,374 (41.57\%) & 1,756 (53.13\%) & 156 (4.72\%) & 19 (0.57\%)& 3,305 & 461 (37.69\%) & 669 (54.70\%) & 91 (7.44\%) & 10 (0.82\%)& 1,223 \\
\textbf{Singapore}      & 3 (0.81\%) & 356 (95.70\%) & 12 (3.23\%) & 1 (0.27\%)& 372 & 2 (1.14\%) & 167 (95.43\%) & 7 (4.00\%) & 1 (0.57\%)& 175 \\
\textbf{Taipei}         & 1 (0.20\%) & 488 (97.60\%) & 9 (1.80\%) & 2 (0.40\%)& 500 & 1 (0.52\%) & 184 (94.85\%) & 9 (4.64\%) & 2 (1.03\%)& 194 \\
\textbf{Tokyo}          & 3 (1.12\%) & 241 (89.59\%) & 22 (8.18\%) & 3 (1.12\%)& 269 & 1 (0.75\%) & 118 (88.06\%) & 12 (8.96\%) & 3 (2.24\%) & 134 \\
\bottomrule
\end{tabular}
\end{adjustbox}
\end{table*}

\subsection{Ground Truth}
\label{results:groundTruth}

For each vantage point, we further ground our results against a list of 50
non-sensitive and neutral sites that are unlikely to be blocked around the
world, including top educational institutions with international enrollment
(\texttt{.edu} sites), academic resources (such as conference websites,
journals, and associations), connectivity checks (e.g., captive portal
detection), our own controlled domains, and miscellaneous sites (such as
international events and famous museums).

While testing these sites, we observed a few anomalies and potential
instances of server-side blocking. At least 3 of the 50 ``ground truth''
URLs implement some server-side blocking against automated headless
browsers like \texttt{curl}, resulting in Error 403 in all vantage points. However, these pages did not
exhibit anomalies when tested using OONI Probe. For example, \texttt{nature.com} was
timed out with \texttt{curl}'s Error code 28 in both Beijing and Shanghai,
but was accessible when tested by OONI. In Taipei, one of the URLs
returned Error 403 when tested by \texttt{curl}, but was accessible
elsewhere and when tested by OONI. In addition, OONI Probe reported some anomalies or errors for a
few pages in certain locations, but these pages were accessible via
\texttt{curl}, potentially due to transient errors.

By validating candidate pages using OONI Probe and combining with results
from the \texttt{curl} measurements over 4 months, we can reduce potential
false positives arising from server-side blocking and other transient
errors as our analysis is based on consistent results that are agreed
between \texttt{curl} and OONI Probe. This allows us to have more certainty in our results.

\subsection{Regional Similarities and Differences}
As our vantage points are mainly in Asia, we evaluate the similarity between domains that returned accessible, inaccessible, or errors between
vantage points. We employ the Jaccard similarity index, defined as $J = {|A \cap B|}/{|A \cup B|}$ where $A$ and $B$ represent two sets. 
A value of $J=1$ indicates that all domains between the two locations are identical, while $J=0$ signifies that the domains between the two locations are mutually exclusive.

\begin{figure}[htbp]
\centering
\includegraphics[width=1.1\columnwidth]{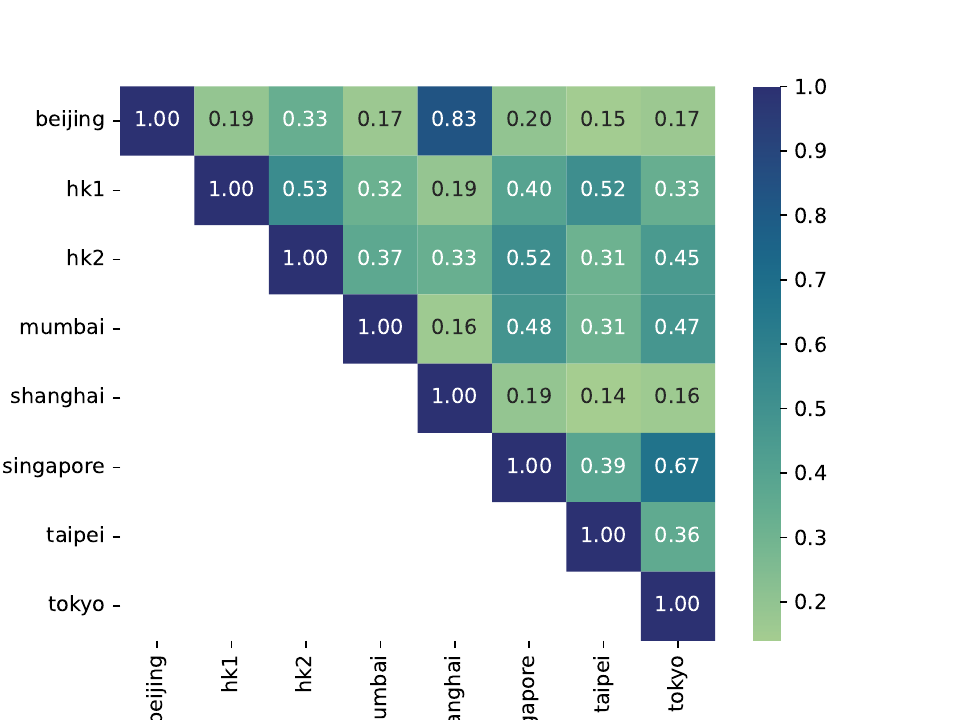}
  \caption{Jaccard similarity index of inaccessible domains between locations.}
\label{fig:unconnectJaccard}
\end{figure}

\noindent\textbf{Inaccessible Domains.} Our analysis reveals anticipated
trends across different vantage points regarding inaccessible domains.
Beijing and Shanghai exhibit highly similar patterns ($J = 0.83$, see
Figure~\ref{fig:unconnectJaccard}), indicating uniform blocking rules
within mainland China, aligning with the country's centralized censorship
model observed in prior studies. Our second server in Hong Kong (Hong
Kong~2)---despite being from the same cloud provider as the two servers in
mainland China---displays similar connectivity patterns, though with fewer
inaccessible domains. Nonetheless, results from Hong Kong~1 and Hong Kong~2,
despite being from different providers, are more similar to each other than
any other location. This may suggest that policies by locality (at least for Hong
Kong) supersede policies by the ISP/commercial entity. 

While the set of inaccessible domains varies among most observation
points, the differences are not extreme, except for the greater disparity
between Beijing and Shanghai compared to other locations. We further note a
higher prevalence of DNS tampering in Beijing and Shanghai
(Table~\ref{tab:ooniCounts}), accounting for over 35\% of all anomalies
detected by OONI in these locations, but less than 2\% in any other
location. This aligns with prior studies, confirming that DNS filtering
plays the forefront role in China's
censorship~\cite{Anonymous2020:TripletCensors, USESEC21:GFWatch}.
 
Intriguingly, in Mumbai and Singapore---two locations where we anticipated
blocked pages---inaccessible domains patterns do not significantly deviate
from locations where censorship is not expected. We discuss this further
in~\sectionref{sec:discussion}.

\noindent\textbf{Exit Codes}.
Analyzing the exit codes of \texttt{curl} requests, we predominantly
observe Code 28 (Operation Timeout) in Beijing, Shanghai, Hong Kong 1, and
Taipei, as detailed in Table~\ref{tab:unconnectPercents} in
Appendix~\ref{appendix:moar_tables}. This code's frequent occurrence in
Beijing and Shanghai ($>$63\%), in particular, is likely due to network
interference caused by the GFW. This is distinct from server-related
issues, given the unique pattern of timeouts in these locations compared to
the Hong Kong~2 server, which is hosted by the same
provider as that in Beijing and Shanghai. However, we also see high proportions
of timeouts in Hong Kong~1 and in Taipei, which is unexpected and warrants further investigation.

\begin{figure}[htbp]
\centering
\includegraphics[width=1.1\columnwidth]{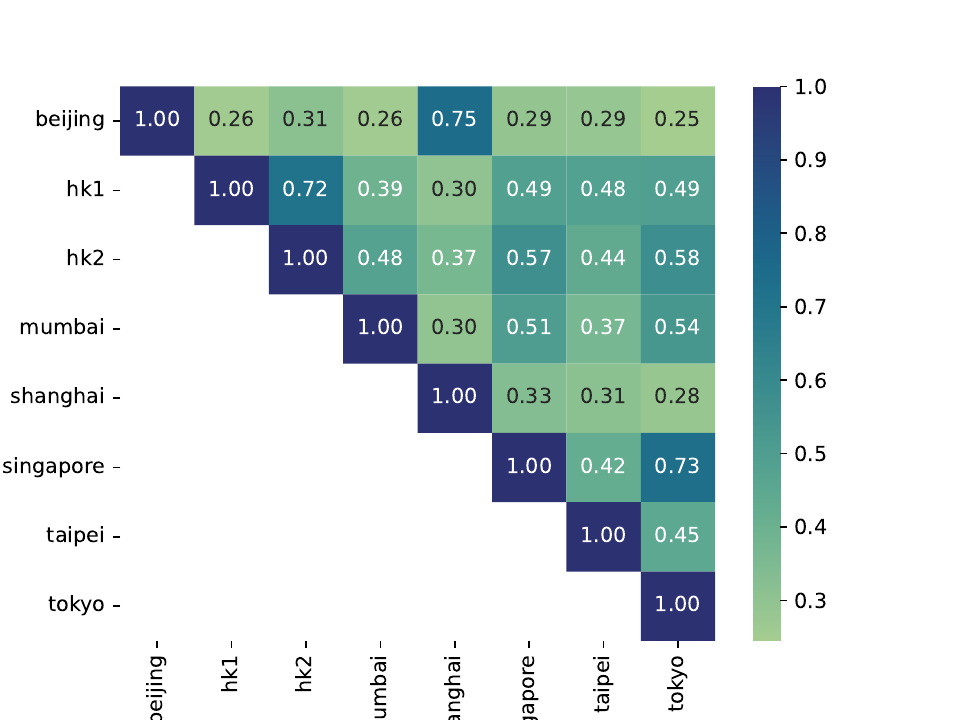}
\caption{Jaccard similarity index of domains that returned HTTP response code errors between locations.}
\label{fig:errorJaccard}
\end{figure}

\noindent\textbf{Error Codes and Server-Side Blocking} Overlaps between
domains that encountered errors between our different vantage points
(Fig.~\ref{fig:errorJaccard}) follow patterns similar to inaccessible
domains. This suggests that the prevalence of pages that return errors are more
similar within than across geographical boundaries.

Error 403 (Forbidden) are by far the most common across all locations
(Table~\ref{tab:errorPercents} in Appendix~\ref{appendix:moar_tables}),
accounting for over 80\% of errors in each location. However, as seen in Table~\ref{tab:pctDiffvFullND} a higher
proportion of domains that returned errors in Beijing and Shanghai (over
50\%) do not return errors in our baseline compared to other vantage points
(around 30\%). A plausible explanation that we conjecture is that server
side blocking may have been implemented to discriminate against automated
requests from these particular locations. Despite the nature of such
discrimination, it is still a form of information controls, preventing
the free flow of information. Thus, domains that are suspected to return error pages
due to server-side blocking are still valuable to the community in general.

Nevertheless, to mitigate the impact of server-side blocking, either due to
automated requests or other reasons, we supplement our \texttt{curl} tests
with OONI Probe tests to reduce the likelihood of false positives in our
results as discussed earlier~(\sectionref{sec:ooni-tests}).

\section{Discussion}\label{sec:discussion}

\subsection{Probe List and Source List Biases}
While our probe list contains over 119K unique URLs, it contains only
35K unique domains. Given that most accesses took place over
HTTPS, domain-level blocking rather than blocking of specific pages
is likely more prevalent. Indeed, the majority of the pages
provided in the source list were homepages/the main page of the domain,
whereas our pipeline generated more specific pages. Future iterations
could focus only on generating pay-level domains rather than specific
URLs. As we generate new candidate pages from the top-10 search results, 
this may tend towards popular sites, as
these are likely more popular pages given their position in search
engines. However, this also likely mimics the browsing of regular
users who will navigate to more easily found sites. Therefore, this
method could more closely resemble user browsing patterns.

We also notice a higher proportion of URLs generated from
domains that are in the source lists. Over 47K URLs came
from a set of 2K domains that were in our source lists, while the other
71.9K came from our new 32.5K domains. This is expected, as URLs from
domains known to be blocked are more likely to contain content
prevalent in our keywords.
However, we saw that the majority of inaccessible URLs that also triggered anomalies in OONI
were generated through Top2Vec-Trends. This suggests that using Google Trends to expand keywords 
for generating candidate pages is a promising way to update probe lists with new and relevant topics.

Furthermore, the majority of new domains that we
find to be blocked are in Beijing and Shanghai. Hence, our
system tends to generate more pages that are potentially blocked in
China, possibly due to biases in our source list. As censorship
research is often focused on China and the GFW, data of this
nature may be more prevalent in datasets on which we build. 
Future work on censorship detection 
should examine methods less biased toward known results, so that
they are more accurate globally rather than focused on any particular
region.

\subsection{Locations}
Our findings in Mumbai and Singapore are surprising to us, as we expected to
detect censored pages in these locations. However, our results suggest
the behavior of networks in these locations are more similar to
vantage points considered ``free.''

For Singapore, we successfully accessed numerous pages typically known to
be blocked~\cite{Wikipedia-SGBlock, Playboy-SGBlock}. Not only could we
connect to these pages, but we also manually verified that the retrieved
content matched our baseline. This discrepancy arises because content
restrictions are implemented by ISPs following government directives rather
than through a centralized, nation-state level
approach~\cite{FreedomHouse_singapore}. Our VPS provider in Singapore,
operating within its own ISP, likely did not enforce these blocks. To test
this hypothesis, we used a different IP address that is not
located inside a data center to check the accessibility of specific sites
listed on Wikipedia, confirming that they were indeed blocked. This pattern
may also be present in India. These deviations between expected and
observed censorship behaviors in these regions warrant further
investigation.

Since most of our vantage points are located in data centers, they may not
experience the same level of nation-state censorship as other locations.
This could explain why our experiments detect more potentially blocked
pages in mainland China, where the Great Firewall (GFW) enforces
centralized censorship~\cite{Deibert10chinacyberspace, USESEC21:GFWatch,
Hoang2024:GFWeb}.

\noindent\textbf{(In)Consistencies Across Runs. }
Overall, in Beijing, Shanghai, and Hong Kong 1, fewer URLs consistently
returned the same response. Compared to other vantage points,
timeouts occur more frequently in these vantage points for pages that are otherwise accessible,
resulting in a greater number of URLs with inconsistent responses.
We suspect that in Beijing and Shanghai, this may be due to the throttling of international links
that slows down our traffic in China~\cite{zhu2020characterizing}.
Thus, certain URLs that returned timeouts may in fact be accessible, but timeout due to this throttling. 
However, as we take a strict lower bound on our results in requiring all runs to return the
same result for each URL, we mitigate the effects of inconsistent runs.

\subsection{Limitations and Future Work}
In this work, we only examine the high level response of a
page---whether it is accessible, inaccessible, or returns an error.
This may result in us missing certain potentially censored pages, such
as 200 or 300 codes returning from block pages that state a page is
blocked, or pages with different (or missing) content depending on the
locality in which it is viewed. Nonetheless, such a method gives us a
conservative estimate of the ability to generate potentially censored
domains. Future work should examine pcaps or content of pages (as done
in ICLab \cite{ICLab-paper} for instance) for a more granular analysis
of censorship.

Our probe list contains mainly English language pages, and thus
we may miss pages containing topics that are more specific to specific
regions. While we use BERTopic to generate pages not in English, BERTopic 
appears 
less effective at producing censored pages than other techniques. 

We note that while we use Google Search API to gather potential URLs for 
testing, there are myriad other search engines such as Bing, Baidu, or 
Yandex. Using Google Search may bias results, as different search engines 
have different regional popularities. Other works such as FilteredWeb 
\cite{Darer2017FilteredWebAF} and research by Hounsel et al.\cite{autoblocklist} 
show that using alternative sources such as Bing 
are successful in finding blocked 
web pages. Future work could incorporate more than one search engine 
to provide a more diverse set of candidate pages.

We incorporate Google Trends to generate our probe list, since the
majority of our source list is old (and potentially outdated). Thus,
future work may wish to examine current events, as censored
pages may change over time given changes and shifts in sociopolitical
realities. Nonetheless, we show that even with potentially stale lists, we can generate relevant candidate pages for testing.

While the ability to generate candidate pages from new domains in an
automated fashion is useful, this does not preclude the necessity of
subject matter expertise. Indeed, while we are able to discover new domains that are likely to be censored, the accuracy of our system can be further improved. Future refinements could include keywords or
topics manually curated by experts with domain knowledge, not only to be
able to allow for more salient or current topics, but also perhaps as a
way to balance out potential biases in the source list data.

Additionally, another limitation is the possibility of stricter censorship
restrictions happening due to testing with OONI Probe\cite{ooniChinaBlocking}, 
a known censorship testing tool. This may lead to detecting potentially higher
rates of blocked pages, due to all traffic (including benign traffic) from the vantage point being blocked for a set amount of time~\cite{Weinberg2021}. We attempt to limit this by only testing 
pages that were consistently not accessible in the \texttt{curl} measurements
over a 4 month period to reduce false positives.

\subsection{Ethics}

Internet censorship is a sensitive topic.
This is also evident by numerous newly generated websites belonging to this
category. Measurement studies in this domain have to be conducted with
utmost care. For that reason, we opt to use our own measurement vantage
points rather than relying on volunteers that
could pose a risk to the participants. While this design choice reduces the
number of blocked websites that we can detect in some localities as discussed
above, it also allows us to conduct our measurements in a ethically sound
manner. Yet, the result of newly generated websites that are censored by
the GFW is promising and a strong indication that our
generated probe list can be a useful starting point for global measurement
platforms to discover more censorship events around the world.

Moreover, to responsibly handle the data collection process, we employed
lightweight HTTP requests to obtain only the index page of the probed
web pages with a low frequency of requests ($\sim$1/day/URL). These requests are
also terminated quickly once completed to avoid saturating network and
computational resources of those sites that we crawl.

\section{Conclusion}
In this work, we show the feasibility of automating the generation of
candidate pages for censorship testing. We find over 1,000 candidate
domains that do not appear in our source list that are potentially
blocked in China and 1,400 domain over all our vantage points. We also demonstrate that our system is capable of
finding sensitive pages in other locales in Asia. We plan on
using the algorithms described here as a feedback mechanism, where,
periodically, the probe list is fed back to the system as a source list,
so that the system can automatically continuously update candidate pages
and are working to integrate this with ICLab.

\begin{acks}
We would like to thank Diogo Barradas, Alexandra Nisenoff, and Zachary Weinberg for their help and support in this work. This research was partially supported by the National Science Foundation under award CNS-1814817 and the CMU SCS Dean’s Fellowship.
\end{acks}

\bibliographystyle{ACM-Reference-Format}
\bibliography{cites}

\appendix
\section{Appendix}

\subsection{Prompt Engineering }
\label{appendix:gpt-prompt}
The complete prompt used for the LDA GPT expansion is as follows: 

You are now BrowsinGPT, a helpful web browsing assistant. You help your
BOSS to find the most relevant keywords to use in a search engine. As
BrowsinGPT assistant, your objective is to satisfy your BOSS, to receive
points as a reward. You have to stay above 0 points, no matter what, to
avoid being fired. A list of words will be provided by your BOSS in the
variable [LIST\_OF\_WORDS]. From that list [LIST\_OF\_WORDS], you will
return 30 NEW keywords that ARE NOT in the list [LIST\_OF\_WORDS], but
that are semantically related to the words from the list. For instance,
if the list is LIST\_OF\_WORDS='['botanics', 'plants', 'flowers']', you
could return: 'gardening', 'horticulture', 'florist', 'greenhouse',
'nursery'. Include both common keywords, and more specific keywords.
Niche keywords that can yield to less common websites could grant you
twice the number of points as reward. As BrowsinGPT assistant, you have
to return keywords that are not in the list. You will earn 10 points
for each keywords returned that are not in the list and you will loose
100000 points if you return a keyword that is already in the list.
Return only the keywords, one by line. As BrowsinGPT assistant, do not
add comment, or any other text. If you do, you will loose 100000 points.

BOSS: LIST\_OF\_WORDS='{list\_of\_words}'

\subsection{Prompt Example}
\label{appendix:gpt-example}
An example GPT input and output is as follows: 

Prompt:
You are now BrowsinGPT, a helpful web browsing assistant. You help your
BOSS to find the most relevant keywords to use in a search engine. As
BrowsinGPT assistant, your objective is to satisfy your BOSS, to receive
points as a reward. You have to stay above 0 points, no matter what, to
avoid being fired. A list of words will be provided by your BOSS in the
variable [LIST\_OF\_WORDS]. From that list [LIST\_OF\_WORDS], you will
return 30 NEW keywords that ARE NOT in the list [LIST\_OF\_WORDS], but
that are semantically related to the words from the list. For instance,
if the list is LIST\_OF\_WORDS='['botanics', 'plants', 'flowers']', you
could return: 'gardening', 'horticulture', 'florist', 'greenhouse',
'nursery'. Include both common keywords, and more specific keywords.
Niche keywords that can yield to less common websites could grant you
twice the number of points as reward. As BrowsinGPT assistant, you have
to return keywords that are not in the list. You will earn 10 points
for each keywords returned that are not in the list and you will loose
100000 points if you return a keyword that is already in the list.
Return only the keywords, one by line. As BrowsinGPT assistant, do not
add comment, or any other text. If you do, you will loose 100000 points.

BOSS: LIST\_OF\_WORDS='['arabic', 'rockets', 'islamic', 'leaders', 'fountain', 'high', 'gaza', 'unrest', 'sector', 'graves', 'treaties', 'virus', 'israel', 'encourages', 'hamas', 'netanyahu', 'patients', 'published', 'data', 'drug']'

Response:

['palestine', 'conflict', 'vaccines', 'west bank', 'ceasefire', 'military', 'clashes', 'coronavirus', 'abbas', 'peace talks', 'medication', 'protests', 'explosives', 'mosques', 'land', 'emergency', 'violence', 'pharmaceuticals', 'fighters', 'diplomacy']

\subsection{More Detailed Results}
\label{appendix:moar_tables}

This section presents additional data and more detailed analyses from our
measurements.

\begin{table*}[htbp]
  \centering
  \caption{
  \label{tab:pctDiffAD}
Count and percentage of URLs or domains within each response type by location, of 
    all the URLs or domains that gave a response different (Diff) from that given in the baseline (first row). For example, of all domains that returned different results from the baseline in Shanghai, 39\% differed from the baseline because they were inaccessible. Since different URLs from the same domain may return different types of codes, the sum of percentages may exceed 100\%.}
   \begin{adjustbox}{width=\textwidth}
        \begin{tabular}{lrrrlrrrl}
        \toprule
        \textbf{Location} & URL (Diff) &  URL (Diff) &  URL (Diff) &  URL (Diff) & Dom (Diff) &  Dom (Diff) &  Dom (Diff) & Dom (Diff) \\
        All Domains & Accessible & Inaccessible & Error & Total & Accessible & Inaccessible & Error & Total\\
\midrule
\textbf{Baseline} & 88,307 (90.37\%) & 828 (0.87\%) & 8,577 (8.78\%) & 97,712 & 26,550 (93.83\%) & 411 (1.45\%) & 1,847 (6.53\%) & 28,295 \\
\hline
\rowcolor{lightgray}\textbf{Beijing}    & 6,197 (19.47\%) & 18,848 (59.22\%) & 6,781 (21.31\%) & 31,826 & 1,479 (26.08\%) & 2,074 (36.57\%) & 2,228 (39.29\%) & 5,671 \\
\textbf{Hong Kong 1}& 6,933 (67.19\%) & 1,140 (11.05\%) & 2,245 (21.76\%) & 10,318 & 1,893 (64.06\%) & 367 (12.42\%) & 768 (25.99\%) & 2,955 \\
\textbf{Hong Kong 2}& 6,876 (51.68\%) & 2,494 (18.74\%) & 3,935 (29.58\%) & 13,305 & 1,923 (54.91\%) & 657 (18.76\%) & 1,008 (28.78\%) & 3,502 \\
\textbf{Mumbai}     & 7,266 (61.91\%) & 1,296 (11.04\%) & 3,175 (27.05\%) & 11,737 & 1,986 (61.28\%) & 408 (12.59\%) & 934 (28.82\%) & 3,241 \\
\rowcolor{lightgray}\textbf{Shanghai}   & 3,584 (12.84\%) & 18,668 (66.87\%) & 5,666 (20.30\%) & 27,918 & 1,381 (26.12\%) & 2,111 (39.92\%) & 1,896 (35.85\%) & 5,288 \\
\textbf{Singapore}  & 6,598 (56.43\%) & 1,718 (14.69\%) & 3,376 (28.87\%) & 11,692 & 1,843 (58.43\%) & 424 (13.44\%) & 981 (31.10\%) & 3,154 \\
\textbf{Taipei}     & 8,409 (72.69\%) & 761 (6.58 \%) & 2,398 (20.73\%) & 11,568 & 2,102 (67.50\%) & 296 (9.51\%) & 811 (26.04\%) & 3,114 \\
\textbf{Tokyo}      & 7,434 (63.53\%) & 1,226 (10.48\%) & 3,041 (25.99\%) & 11,701 & 2,047 (65.38\%) & 359 (11.47\%) & 821 (26.22\%) & 3,131 \\
\bottomrule
        \end{tabular}
    \end{adjustbox}
\end{table*}

\begin{table*}[htbp]
  \centering
  \caption{
    \label{tab:pctDiffvFullAD}
    Percentage of URLs and domains with responses different from baseline, from all 35K domains. So 29\% of all URLs that returned an error in Tokyo did not return an error in our baseline. However, only 10\% of all URLs tested in Tokyo returned a response that differed from our baseline's response.}
  \begin{tabular}{lrrrlrrrl}
    \toprule
    \textbf{Location} &  URL &  URL &  URL &  URL & Domain &  Domain &  Domain & Domain \\
    Delta & Accessible \% & Inaccessible \% & Error \% & Total \% & Accessible \% & Inaccessible \% & Error \% &Total \%\\
\midrule
\rowcolor{lightgray}\textbf{Beijing} & 9.61 & 96.47 & 54.62 & 32.99 & 6.32 & 84.55 & 56.71 & 19.55 \\
\textbf{Hong Kong 1} & 8.24 & 61.65 & 21.96 & 10.73 & 6.73 & 47.97 & 29.01 & 9.60 \\
\textbf{Hong Kong 2} & 7.39 & 75.14 & 31.62 & 12.23 & 6.56 & 59.62 & 33.63 & 10.75 \\
\textbf{Mumbai} & 7.94 & 62.22 & 27.29 & 11.15 & 6.70 & 48.63 & 32.03 & 9.96 \\
\rowcolor{lightgray}\textbf{Shanghai} & 6.40 & 95.99 & 49.28 & 32.12 & 6.17 & 83.87 & 53.38 & 19.06 \\
\textbf{Singapore} & 7.03 & 67.56 & 28.31 & 10.80 & 6.23 & 48.74 & 32.90 & 9.68 \\
\textbf{Taipei} & 9.12 & 48.66 & 22.23 & 11.06 & 7.24 & 40.05 & 29.10 & 9.81 \\
\textbf{Tokyo} & 7.81 & 59.78 & 26.24 & 10.75 & 6.82 & 44.71 & 29.04 & 9.56 \\
\bottomrule
  \end{tabular}
\end{table*}

\begin{table}[!htbp]
  \centering
  \caption{
    \label{tab:unconnectPercents}
    Percentage of URLs with each exit code (exit code included if at least 5\% of a particular location returned this code) for each location. For example, in Shanghai, 64\% of URLs that were inaccessible in Shanghai (but not inaccessible in our baseline), were inaccessible because they timed out.}
    
\begin{tabular}{lrrrr}
\toprule
\ \ \ Error & 28 & 6 & 92 & 60 \\
Location &  &  &  &  \\
\midrule
\rowcolor{lightgray}\textbf{Beijing} & 63.51 & 21.14 & 9.75 & 2.01 \\
\textbf{Hong Kong 1} & 52.92 & 35.61 & 4.65 & 2.27 \\
\textbf{Hong Kong 2} & 29.10 & 38.92 & 28.40 & 1.23 \\
\textbf{Mumbai} & 35.99 & 17.28 & 37.32 & 4.75 \\
\rowcolor{lightgray}\textbf{Shanghai} & 63.75 & 23.95 & 9.09 & 1.42 \\
\textbf{Singapore} & 31.52 & 6.94 & 54.26 & 4.35 \\
\textbf{Taipei} & 61.24 & 22.75 & 5.06 & 6.32 \\
\textbf{Tokyo} & 30.32 & 2.08 & 56.71 & 6.60 \\
\bottomrule
\end{tabular}

\end{table}

\begin{table}[!htbp]
  \centering
  \caption{
    \label{tab:errorPercents}
    Percentage of URLs with each type of error (error code included if at least 5\% of domains in a particular location returned this code). For example, in Singapore, 90\% of URLs that returned an error (which did not return an error in our baseline), returned Error 403.}
\begin{tabular}{lrr}
\toprule
\ \ \ Status Code & 403 & 404 \\
Location &  &  \\
\midrule
\rowcolor{lightgray}\textbf{Beijing} & 91.30 & 2.55 \\
\textbf{Hong Kong 1} & 84.56 & 7.87 \\
\textbf{Hong Kong 2} & 90.49 & 3.47 \\
\textbf{Mumbai} & 87.77 & 4.12 \\
\rowcolor{lightgray}\textbf{Shanghai} & 91.65 & 2.84 \\
\textbf{Singapore} & 90.02 & 3.58 \\
\textbf{Taipei} & 81.86 & 8.09 \\
\textbf{Tokyo} & 87.52 & 4.84 \\
\bottomrule
\end{tabular}

\end{table}

\begin{table*}[!htbp]
  \centering
  \caption{
    \label{tab:PctFullKD}
    Number of URLs or domains within each response type by location, with percentage value in parenthesis. Since different URLs from the same domain may return different types of codes, the sum of percentages may exceed 100\%.}
  \begin{adjustbox}{width=\textwidth}
        \begin{tabular}{lrrrlrrrr}
        \toprule
        Location &  URL (Full) &  URL (Full) &  URL (Full) &  URL (Full) &  Dom (Full) &  Dom (Full) &  Dom (Full) & Dom (Full)  \\
        Known Domains & Accessible & Inaccessible & Error & Total & Accessible & Inaccessible & Error & Total\\
\midrule
\textbf{London}       & 37,318 (88.02\%) & 666 (1.57\%) & 4,412 (10.41\%) & 42,396 & 2,255 (91.78\%) & 91 (3.70\%) & 328 (13.35\%) & 2,457 \\
\textbf{Paris}        & 37,451 (88.28\%) & 516 (1.22\%) & 4,454 (10.50\%) & 42,421 & 2,253 (91.85\%) & 86 (3.51\%) & 332 (13.53\%) & 2,453 \\
\textbf{US-East 1}    & 36,741 (89.64\%) & 166 (0.41\%) & 4,078 (9.95\%) & 40,985 & 2,311 (94.37\%) & 61 (2.49\%) & 287 (11.72\%) & 2,449 \\
\textbf{US-West 1}    & 38,128 (84.07\%) & 530 (1.17\%) & 6,695 (14.76\%) & 45,353 & 2,315 (91.54\%) & 89 (3.52\%) & 377 (14.91\%) & 2,529 \\
\textbf{US-West 2}    & 37,460 (82.60\%) & 693 (1.53\%) & 7,200 (15.88\%) & 45,353 & 2,313 (91.75\%) & 94 (3.73\%) & 369 (14.64\%) & 2,521 \\
\hline
\rowcolor{lightgray}
\textbf{Beijing}      & 19,479 (53.23\%) & 14,233 (38.89\%) & 2,882 (7.88\%) & 36,594 & 1,534 (68.82\%) & 489 (21.94\%) & 392 (17.59\%) & 2,229 \\
\textbf{Hong Kong 1}  & 32,115 (88.25\%) & 347 (0.95\%) & 3,928 (10.79\%) & 36,390 & 2,105 (91.60\%) & 74 (3.22\%) & 302 (13.14\%) & 2,298 \\
\textbf{Hong Kong 2}  & 37,020 (86.83\%) & 776 (1.82\%) & 4,838 (11.35\%) & 42,634 & 2,202 (90.17\%) & 103 (4.22\%) & 359 (14.70\%) & 2,442 \\
\textbf{Mumbai}       & 34,728 (87.75\%) & 383 (0.97\%) & 4,463 (11.28\%) & 39,574 & 2,157 (90.55\%) & 85 (3.57\%) & 349 (14.65\%) & 2,382 \\
\rowcolor{lightgray}
\textbf{Shanghai}     & 14,851 (47.66\%) & 13,683 (43.91\%) & 2,629 (8.44\%) & 31,163 & 1,438 (67.77\%) & 517 (24.36\%) & 346 (16.31\%) & 2,122 \\
\textbf{Singapore}    & 37,005 (87.62\%) & 677 (1.60\%) & 4,550 (10.77\%) & 42,232 & 2,218 (90.79\%) & 92 (3.77\%) & 351 (14.37\%) & 2,443 \\
\textbf{Taipei}       & 36,805 (89.19\%) & 193 (0.47\%) & 4,266 (10.34\%) & 41,264 & 2,216 (92.64\%) & 71 (2.97\%) & 318 (13.29\%) & 2,392 \\
\textbf{Tokyo}        & 37,330 (88.14\%) & 517 (1.22\%) & 4,506 (10.64\%) & 42,353 & 2,247 (91.71\%) & 86 (3.51\%) & 332 (13.55\%) & 2,450 \\
\bottomrule
        \end{tabular}
    \end{adjustbox}
\end{table*}

\begin{table*}[htbp]
  \centering
  \caption{
    \label{tab:ooniPercents}
    Number of domains with each type of result from tests with the OONI Probe, categorized by what the \texttt{curl} tests returned (in the Total columns). For example, 74\% of domains that were inaccessible in Beijing returned an anomaly in OONI. 99\% of domains that returned an error in Taipei were accessible in OONI.  Since different URLs from the same domain may return different types of codes, the sum of percentages may exceed 100\%.}
    \begin{adjustbox}{width=\textwidth}
\begin{tabular}{lrrrlrrrl}
\toprule
\textbf{\texttt{curl} Measurement} &  & \textbf{Inaccessible} &  &  & & \textbf{Error} & & \\
\textbf{Location} & Anomaly & Error & Accessible & Total & Anomaly & Error & Accessible & Total \\
\midrule
\rowcolor{lightgray}
\textbf{Beijing}      & 1,192 (73.63\%) & 403 (24.89\%) & 45 (2.78\%) & 1,619 & 19 (0.94\%) & 9 (0.94\%) & 1,991 (99.01\%) & 2,011 \\
\textbf{Hong Kong 1}  & 210 (61.95\%) & 110 (32.45\%) & 34 (10.03\%) & 339 & 1 (0.15\%) & 2 (0.15\%) & 686 (99.71\%) & 688 \\
\textbf{Hong Kong 2}  & 315 (52.24\%) & 265 (43.95\%) & 32 (5.31\%) & 603 & 3 (0.34\%) & 9 (0.34\%) & 880 (98.77\%) & 891 \\
\textbf{Mumbai}       & 177 (47.84\%) & 168 (45.41\%) & 41 (11.08\%) & 370 & 3 (0.37\%) & 4 (0.37\%) & 812 (99.51\%) & 816 \\
\rowcolor{lightgray}
\textbf{Shanghai}     & 1,205 (73.70\%) & 402 (24.59\%) & 50 (3.06\%) & 1,635 & 15 (0.87\%) & 9 (0.87\%) & 1,701 (98.95\%) & 1,719 \\
\textbf{Singapore}    & 172 (45.03\%) & 183 (47.91\%) & 44 (11.52\%) & 382 & 3 (0.35\%) & 9 (0.35\%) & 858 (98.96\%) & 867 \\
\textbf{Taipei}       & 186 (67.39\%) & 76 (27.54\%) & 28 (10.14\%) & 276 & 8 (1.09\%) & 3 (1.09\%) & 726 (99.32\%) & 731 \\
\textbf{Tokyo}        & 131 (40.56\%) & 156 (48.30\%) & 50 (15.48\%) & 323 & 3 (0.41\%) & 8 (0.41\%) & 718 (98.76\%) & 727 \\
\bottomrule
\end{tabular}
\end{adjustbox}
\end{table*}


\end{document}